\documentclass[aps,prl,twocolumn,superscriptaddress,showpacs]{revtex4-1}
\usepackage{amsmath,amssymb,graphics,epsfig,epstopdf,color,verbatim,ulem,braket,tabularx}
\usepackage[colorlinks,linkcolor=blue,citecolor=blue,urlcolor=blue]{hyperref}
\newcommand{\beq}{\begin{equation}}
\newcommand{\eeq}{\end{equation}}
\newcommand{\beqn}{\begin{eqnarray}}
\newcommand{\eeqn}{\end{eqnarray}}

\begin{document}

\title{Quantum critical point of Dirac fermion mass generation without spontaneous symmetry breaking}

\author{Yuan-Yao He}
\author{Han-Qing Wu}
\address{Department of Physics, Renmin University of China, Beijing 100872, China}
\author{Yi-Zhuang You}
\affiliation{Department of Physics, University of California,
Santa Barbara, California 93106, USA}
\author{Cenke Xu}
\affiliation{Department of Physics, University of California,
Santa Barbara, California 93106, USA}
\author{Zi Yang Meng}
\affiliation{Beijing National Laboratory
for Condensed Matter Physics, and Institute of Physics, Chinese
Academy of Sciences, Beijing 100190, China}
\author{Zhong-Yi Lu}
\address{Department of Physics, Renmin University of China, Beijing 100872, China}

\begin{abstract}

We study a lattice model of interacting Dirac fermions in $(2+1)$ dimension space-time with an SU(4) symmetry. While increasing interaction strength, this model undergoes a {\it continuous} quantum phase transition from the weakly interacting Dirac semimetal to a fully gapped and nondegenerate phase without condensing any Dirac fermion bilinear mass operator. This unusual mechanism for mass generation is consistent with recent studies of interacting topological insulators/superconductors, and also consistent with recent progresses in lattice QCD community.

\end{abstract}


\date{\today}
\maketitle

\textit{Introduction}. In the Standard Model of particle physics,
all the matter fields, quarks and leptons, acquire their mass from
``spontaneous symmetry breaking", or equivalently the condensation
of the Higgs field~\cite{Higgs1964,Englert1964,Guralnik1964}. The
Higgs field couples to the bilinear mass operator of the Dirac
fermion matter fields (except for the neutrinos), and hence the
matters acquire a mass in the condensate. In the context of
correlated electron systems, mass generation (or gap opening) due
to interaction is also often a consequence of spontaneous symmetry
breaking and the development of certain long-range
order.  
For example, in a
superconductor the Cooper pairs condense, which spontaneously
breaks the $U(1)$ charge symmetry of the electrons, and as a
result the electrons acquire a mass gap at the Fermi
surface. 
So, consensus has that, in strongly interacting fermionic systems
(either in condensed matter or high energy physics), mass (or gap)
generation is usually related to spontaneous symmetry breaking and
the condensation of a fermion bilinear operator~\cite{Landau50}.

However, in condensed matter systems there exists an alternative
mechanism for mass generation, which does not involve any
spontaneous symmetry breaking or long range order. The most
well-known example is the fractional quantum Hall state, where a
partially filled Landau level, which would be gapless without
interaction, is driven into a fully gapped state by strong
interaction. This gapped state has an unusual topological order
and topological ground state
degeneracy~\cite{wen1990,wenniu1990}. Recently, it was discovered
that the phenomenon of ``mass generation without symmetry
breaking" can happen even without topological order. This
mechanism was discovered in the context of interacting topological
insulators, it was found that some topological
insulators/superconductors can be trivialized by interaction. Or
in other words their boundary states, which without interaction
are gapless Dirac fermions or Majorana fermions at one lower
dimension, can be completely gapped out by interaction without
topological degeneracy or condensing any fermion bilinear mass
operator~\cite{fidkowski1,fidkowski2,qiz8,yaoz8,zhangz8,levinguz8,chenhe3B,senthilhe3,xu16,youinversion,Morimoto2015,He2015a,Queiroz2016}.

This new mechanism of mass generation was tested and confirmed
numerically by both condensed matter~\cite{Slagle2015} and lattice
QCD~\cite{Ayyar2015a,Catterall2015,Ayyar2015b} physicists, using
quantum Monte Carlo simulation methods. These works provide evidence that the
massless Dirac fermion phase and the massive quantum phase without
any fermion mass condensation are connected by a single continuous
quantum phase transition.

In this Letter, we construct a microscopic model in $(2+1)$
dimension (D) with four flavors of complex fermions, by employing
large-scale quantum Monte Carlo (QMC) simulations in an unbiased
manner. We find that there indeed exists a single
interaction-driven Dirac semimetal (DSM) to featureless Mott
insulator (FMI) phase transition, which is continuous and does not
involve any spontaneous symmetry breaking. We also provide
analysis of scaling behavior at this novel quantum critical point.

\textit{Model and Method}. We construct a model Hamiltonian with
four-flavors of fermion on a 2D honeycomb lattice at half-filling
with $SU(4)$ symmetry:
\begin{equation}
\label{eq:SU4Model}
\begin{split}
\hat{H}&=H_\text{band} + H_\text{int}\\
\hat{H}_{\text{band}} &= -t\sum_{\langle l,r \rangle\alpha}(-1)^{\alpha}(c_{l\alpha}^{\dagger}c_{r\alpha}+ c_{r\alpha}^{\dagger}c_{l\alpha}) \\
\hat{H}_{\text{int}} &= V\sum_{r}(c_{r1}^{\dagger}c_{r2}c_{r3}^{\dagger}c_{r4} + c_{r4}^{\dagger}c_{r3}c_{r2}^{\dagger}c_{r1}),
\end{split}
\end{equation}
where $\alpha=1,2,3,4$ in $\hat{H}_{\text{band}}$ stands for
fermion flavors and $\langle l,r\rangle$ denotes the
nearest-neighbor sites.
\begin{figure}[h!]
\centering
\includegraphics[width=0.9\columnwidth]{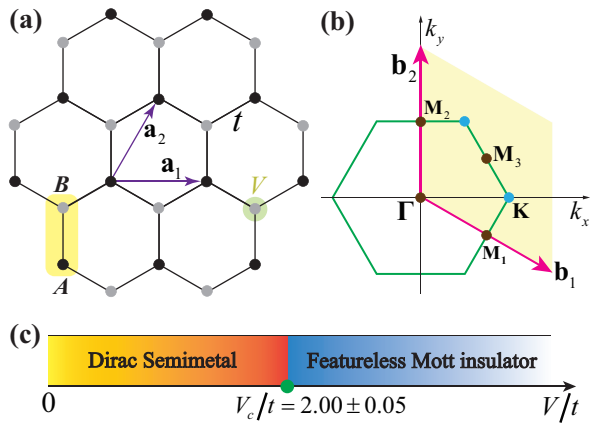}
\caption{\label{fig:LatticeDiagram}(color online) Lattice geometry
and phase diagram for the $SU(4)$ symmetric model in
Eq.~(\ref{eq:SU4Model}). (a) The honeycomb lattice, whose unit cell
is denoted by the yellow shaded rectangle. (b) The Brillouin zone. (c) Phase diagram for the model
Eq.~(\ref{eq:SU4Model}) obtained from QMC simulations. Two quantum
phases, Dirac semimetal and featureless Mott insulator, are
observed, which are connected by a continuous quantum phase
transition located at $V_c/t = 2.00\pm 0.05$.}
\end{figure}
$t$ is set as the energy unit throughout this Letter. The
lattice geometry and Brillouin zone are shown in
Fig.~\ref{fig:LatticeDiagram} (a) and (b), respectively. This Hamiltonian has an
$SU(4)$ symmetry and is invariant under the transformation
$\xi_r\to U\xi_r$ for any $U\in SU(4)$, with
$\xi_r=(c_{r1}^{\dagger},c_{r2},c_{r3}^{\dagger},c_{r4})^{\text{T}}$.
The $(-1)^{\alpha}$ factor in the hopping term
$\hat{H}_{\text{band}}$ is enforced by the $SU(4)$ symmetry.

It is straightforward to check that, if we keep the system at
half-filling, then analogous to the usual case in graphene, all
the lattice symmetries, such as $60^{\circ}$ rotation, reflection,
translation, time-reversal, etc, together with the $SU(4)$ flavor
symmetry and particle-hole symmetry $c_{r\alpha}\to (-1)^r
c_{r\alpha}^\dagger$ prohibit the gap opening of the Dirac
fermions in the noninteracting limit, namely any fermion bilinear
mass operator of the Dirac fermion will break at least one of the
symmetries.


To explore the ground state properties of the model in
Eq.~(\ref{eq:SU4Model}) in the presence of interaction, we employ
projector determinantal quantum Monte Carlo
method~\cite{AssaadEvertz2008,Meng2010}, details of this
calculation are presented in Sec. I of the supplemental
material~\cite{Suppl}. As discussed there, QMC is immune from
minus-sign-problem for both $V>0$ and $V<0$ cases. Comparisons
between exact diagonalization and QMC simulations on a $2\times2$
system (8 lattice sites) are carried out for sanity check.
Numerical verification of the $SU(4)$ symmetry of the model is
also performed and presented in Sec. IV of supplemental
material~\cite{Suppl}. In this Letter, we focus on the $V>0$ case
and the system sizes simulated are $L=3,6,9,12,15,18$. We denote
$N_s=2L^2$ as the total number of lattice sites and $N=L^2$ as
number of unit cells.

\textit{Ground state phase diagram}. The phase diagram of the
$SU(4)$ symmetric model in Eq.~(\ref{eq:SU4Model}) is presented in
Fig.~\ref{fig:LatticeDiagram}(c). Two quantum phases, a gapless
Dirac Semimetal and a featureless Mott insulator, are
observed respectively. Furthermore, they are connected by a continuous quantum
phase transition located at $V_c/t = 2.00\pm0.05$. While
increasing interaction strength $V/t$, we observe no spontaneous
symmetry breaking. The FMI is gapped in both fermionic and bosonic
channels (shown later) without any symmetry breaking.

The FMI is easy to understand from the $V\to+\infty$ limit. Since
the interaction is on-site, it is easy to perceive that, when
$V\to+\infty$, the ground state is
\begin{equation}
\label{eq:SU4ModelVle0} |\Psi_g\rangle =
\prod_r|\Psi_r\rangle=\prod_r\frac{1}{\sqrt{2}}\left(
\prod_{\alpha=1}^4 \xi^\dagger_{r,\alpha} - 1\right)
|0\rangle_{\xi},
\end{equation}
where $|0\rangle_\xi$ is the vacuum of $\xi$ fermions, and
$\hat{H}_{\text{int}}|\Psi_g\rangle = - VN_{s}|\Psi_g\rangle$
(this state is at half-filling written with the $c_{r,\alpha}$
fermions). $|\Psi_g\rangle$ is a direct product state of $SU(4)$
singlets~\cite{He2015a,He2015b,He2015c,You2015}. Since obviously
$|\Psi_g\rangle$ preserves all the symmetries (including flavor,
lattice, time-reversal and particle-hole symmetries) of the
system, any Dirac fermion mass operator should have zero
expectation value in this state. Hence, the wave function
$|\Psi_g\rangle$ describes a symmetric featureless Mott insulator.
Note our state has a different flavor symmetry and number of
states per site compared with another featureless Mott insulator
proposed recently~\cite{Chen2016}.


It is well-known that the (2+1)D massless Dirac fermions are
stable against weak short range interactions~\cite{Meng2010}. The
transition from the weakly interacting DSM to the strongly coupled
FMI as a function of $V/t$ is the main issue that we explore in
this Letter. As it will become clear in the following, a direct
continuous quantum phase transition from DSM to FMI is revealed by
our QMC simulations. More importantly, there is no spontaneous
symmetry breaking and no fermion bilinear condensation across this
transition.

\begin{figure}[h!]
\centering
\includegraphics[width=0.9\columnwidth]{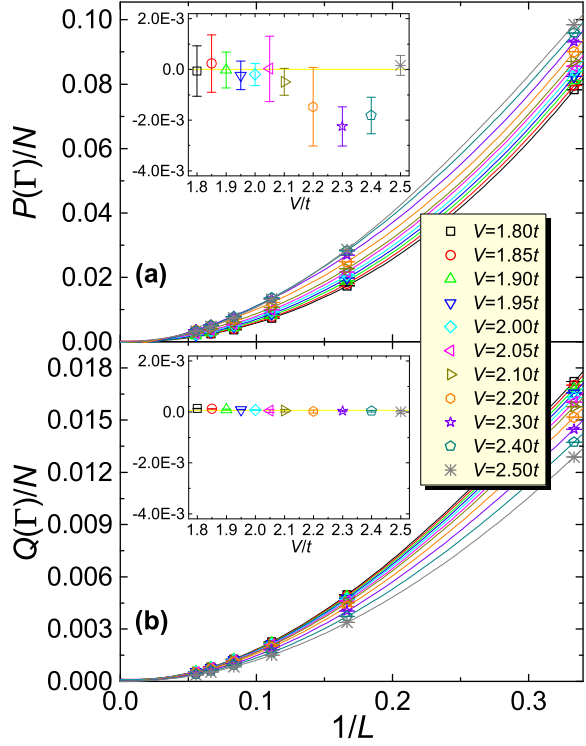}
\caption{\label{fig:O6Extrapolation}(color online) Extrapolation
of structure factors (a) $P(\boldsymbol{\Gamma})/N$ and (b)
$Q(\boldsymbol{\Gamma})/N$ over the inverse system size $1/L$ by
cubic polynomials. The insets show the extrapolated values at the thermodynamic limit. From the results, both of the $O(6)$ orders are absent across the DSM-FMI phase transition.}
\end{figure}

\textit{$O(6)$ order vectors and excitation gaps}. To verify our conclusion, we need to analyze the behavior of all
the Dirac fermion mass operators. Because there is only on-site
interaction in our model, we will focus on Dirac mass operators that are defined on-site, which is most
likely favored by the interaction at the mean field level. We
begin with order parameters that transform as a vector under $SU(4)$ symmetry. Such order parameters can be combined into two
sets of $SO(6)\sim SU(4)$ vector $\boldsymbol{\phi}$ and
pseudo-vector $\boldsymbol{\psi}$~\cite{Suppl}:
\begin{eqnarray}
\phi_{r1} + i \psi_{r1} &=& (c_{r1}^{\dagger}c_{r4}+c_{r3}^{\dagger}c_{r2}), \nonumber \\
\phi_{r2} + i \psi_{r2} &=& (c_{r1}^{\dagger}c_{r3}^{\dagger}+c_{r2}c_{r4}), \nonumber \\
\phi_{r3} + i \psi_{r3} &=& (c_{r1}^{\dagger}c_{r2}-c_{r3}^{\dagger}c_{r4}), \nonumber \\
\phi_{r4} + i \psi_{r4} &=& i(c_{r1}^{\dagger}c_{r4}-c_{r3}^{\dagger}c_{r2}), \nonumber \\
\phi_{r5} + i \psi_{r5} &=& i(c_{r1}^{\dagger}c_{r3}^{\dagger}-c_{r2}c_{r4}), \nonumber \\
\phi_{r6} + i \psi_{r6} &=& i(c_{r1}^{\dagger}c_{r2}+c_{r3}^{\dagger}c_{r4}),
\end{eqnarray}
and the $SO(6)$ symmetry rotates the six components to one
another, respectively. The fact that $\boldsymbol{\phi}$ and
$\boldsymbol{\psi}$ are mass operators of the Dirac fermions is
more explicitly in the basis of $\xi_{r}$ fermions. In the
long-wave-length limit, we can express $\xi_{r}$ in terms of the
low-energy modes $\xi_{K}$($\xi_{K'}$) around the $K$($K'$) point
in the Brillouin zone, as $\xi_{\mathbf{r}} \sim
\xi_{K}e^{i\mathbf{K}\cdot\mathbf{r}}+\xi_{K'}e^{-i\mathbf{K}\cdot\mathbf{r}}$.
The low-energy effective band Hamiltonian reads
\begin{equation}
\begin{split}
H_\text{band}\simeq\int d^2\mathbf{x}\;&\xi_{K}^\dagger v_F(+i\partial_x\sigma^x+i\partial_y\sigma^y)\xi_{K}\\
+&\xi_{K'}^\dagger v_F(-i\partial_x\sigma^x+i\partial_y\sigma^y)\xi_{K'}.
\end{split}
\end{equation}
The operators $\boldsymbol{\phi}+i\boldsymbol{\psi}$ are $SU(4)$
flavor-mixing pairings of the $\xi_{r}$ fermions, which takes the
form of $M_{\alpha\beta}\xi_{K,\alpha}\xi_{K',\beta}$
($\alpha,\beta=1,2,3,4$ label the flavors) with $M$ being a (full
rank) $4\times4$ anti-symmetric matrix. The six orthogonal basis
of the $4\times4$ anti-symmetric matrices correspond to the six
components in $\boldsymbol{\phi}+i\boldsymbol{\psi}$. It is easy
to see that $\boldsymbol{\phi}+i\boldsymbol{\psi}$ can gap out the
Dirac fermions, which are potentially favored to order at the mean
field level.

Due to the $SU(4)$ symmetry, the correlation functions $\langle
\phi_{r,\alpha} \phi_{r^\prime,\alpha} \rangle$ must be identical
for all $\alpha$. The same condition holds for
$\boldsymbol{\psi}$. This is numerically checked and shown in Sec.
IV of supplemental
material~\cite{Suppl}. 

To determine whether the system develops long-range orders in
$\boldsymbol{\phi}$ and $\boldsymbol{\psi}$ with increasing $V/t$,
we measure their structure factors as follows,
\begin{eqnarray}
P(\mathbf{k})&=&\frac{1}{12N}\sum_{\gamma=A,B}\sum_{\eta=1}^6\sum_{ij}e^{i\mathbf{k}\cdot(\mathbf{R}_i-\mathbf{R}_j)} \langle \phi_{i\gamma,\eta}\phi_{j\gamma,\eta}\rangle \nonumber \\
Q(\mathbf{k})&=&\frac{1}{12N}\sum_{\gamma=A,B}\sum_{\eta=1}^6\sum_{ij}e^{i\mathbf{k}\cdot(\mathbf{R}_i-\mathbf{R}_j)} \langle \psi_{i\gamma,\eta}\psi_{j\gamma,\eta}\rangle , \hspace{0.5cm}
\end{eqnarray}
where $i,j$ label unit cells.
Through the extrapolation of $P(\boldsymbol{\Gamma})/N$ and
$Q(\boldsymbol{\Gamma})/N$ over inverse system size $1/L$, we can
obtain the value of $\langle \boldsymbol{\phi} \rangle$ and
$\langle\boldsymbol{\psi}\rangle$ in the thermodynamic limit. The
results for $V/t=1.8 \sim 2.5$ across the phase transition are
shown in Fig.~\ref{fig:O6Extrapolation} (a) and (b), and insets
are the extrapolated values. We notice that the
$Q(\boldsymbol{\Gamma})/N$ is one order of magnitude smaller than
$P(\boldsymbol{\Gamma})/N$.
Combining the results of $P(\boldsymbol{\Gamma})/N$ and
$Q(\boldsymbol{\Gamma})/N$, we conclude that neither
$\boldsymbol{\phi}_l$ nor $\boldsymbol{\psi}_l$ develops
long-range order.


\begin{figure}[tp]
\centering
\includegraphics[width=0.9\columnwidth]{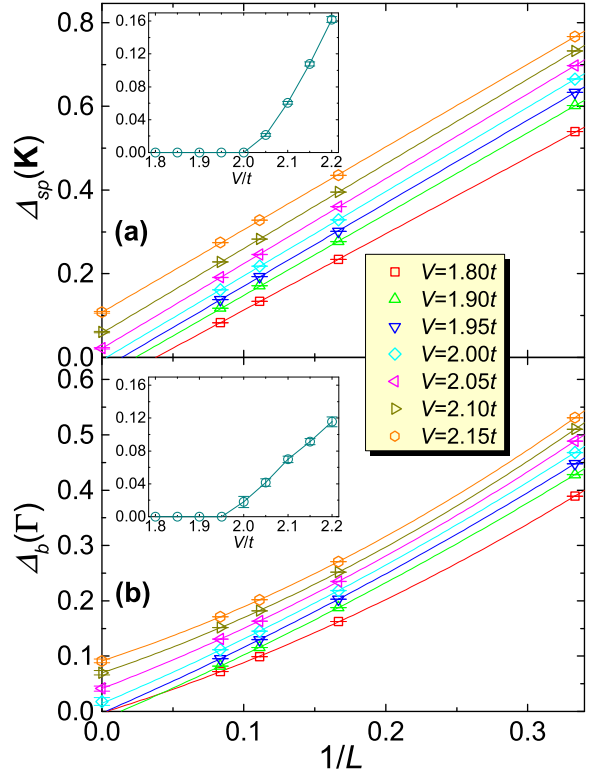}
\caption{\label{fig:GapsExtrapolation}(color online) Extrapolation
of (a) single-particle (fermionic) gap $\Delta_{sp}(\mathbf{K})$
and (b) O(6) order correlation (bosonic) gap
$\Delta_{b}(\boldsymbol{\Gamma})$ over the inverse system size
$1/L$ by linear and quadratic polynomials, respectively. The
insets show the extrapolated gap values at the thermodynamic
limit. Both excitation gaps open at $V_c/t= 2.00\pm0.05$.}
\end{figure}

As for the dynamic properties, the single-particle (fermion) gap
can be extracted from dynamic single-particle Green's function as,
\begin{eqnarray}
G(\mathbf{k},\tau)=\frac{1}{8N}\sum_{\gamma=A,B}\sum_{\alpha=1}^4\sum_{ij}
e^{i\mathbf{k}\cdot(\mathbf{R}_i-\mathbf{R}_j)}[G(\tau)]_{i\gamma,j\gamma}^{\alpha},
\end{eqnarray}
where $[G(\tau)]_{i\gamma,j\gamma}^{\alpha}=\langle T_{\tau}
[c_{i\gamma,\alpha}(\tau)c_{j\gamma,\alpha}^{\dagger}(0)]\rangle$.
The Green's function scales as $G(\mathbf{k},\tau)\propto
e^{-\Delta_{sp}(\mathbf{k})\tau}$ under the limit $\tau\to\infty$
and $\Delta_{sp}(\mathbf{k})$ is the single-particle gap.
Similarly, the bosonic gap $\Delta_b(\boldsymbol{\Gamma})$ can be
extracted from the following dynamic correlation as,
\begin{eqnarray}
P(\mathbf{k},\tau)=\frac{1}{12N}\sum_{\gamma=A,B}\sum_{\eta=1}^6\sum_{ij}e^{i\mathbf{k}\cdot(\mathbf{R}_i-\mathbf{R}_j)}
[P(\tau)]_{i\gamma,j\gamma}^{\eta}, \hspace{0.5cm}
\end{eqnarray}
where $[P(\tau)]_{i\gamma,j\gamma}^{\eta}=\langle T_{\tau}
[\phi_{i\gamma,\eta}(\tau)\phi_{j\gamma,\eta}(0)]\rangle$. Note
that the bosonic gaps extracted from $\boldsymbol{\phi}_l$
correlation and $\boldsymbol{\psi}_l$ correlation should be equal,
which has also been numerically confirmed (see suplemental
material Sec. IV~\cite{Suppl}). Both results of the
single-particle gap and the bosonic gap are shown in
Fig.~\ref{fig:GapsExtrapolation}. Through the extrapolation of the
gap, we observe that the single-particle gap opens at $V/t=2.0
\sim 2.05$, while the bosonic gap opens at $V/t=1.95 \sim 2.0$.
This tiny difference between the critical points extracted from
fermionic and bosonic gap is attributed to finite-size effect, and
the possibility of an intermediate phase with either
$\boldsymbol{\phi}_r$ or $\boldsymbol{\psi}_r$ long-range order
can be ruled out, as otherwise, the single-particle gap should
open before the bosonic gap while increasing $V$. Combining all
data above, we conclude that the DSM-FMI phase transition occurs
at $V_c/t= 2.00\pm0.05$.

\begin{figure}[t]
\centering
\includegraphics[width=0.9\columnwidth]{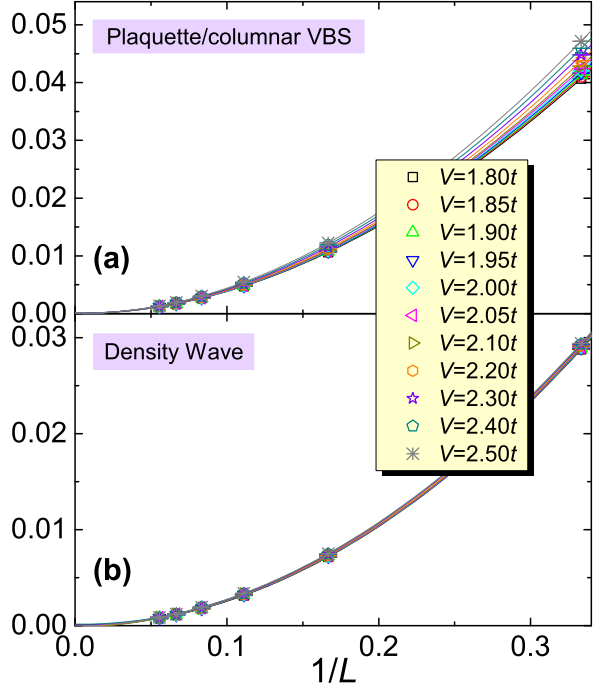}
\caption{\label{fig:OtherOrders}(color online) Extrapolation of
structure factors divided by $N$ for (a) plaquette/columnar VBS
order, (b) density wave order, over inverse system size $1/L$ by
cubic polynomials, across the DSM-FMI phase transition. The
results show that neither of these two long-range orders exists
near the DSM-FMI phase transition.}
\end{figure}

\textit{Other possible long-range orders}. In addition to the two sets of $O(6)$ order parameters, there are
other Dirac fermion mass operators (or order parameters) which may
develop long-range order due to the interaction in
Eq.~(\ref{eq:SU4Model}).
All the possible Dirac mass operators are summarized in
supplemental material Sec. III~\cite{Suppl}. The results of four
representative order parameters, including the plaquette/columnar
valence bond solid (VBS) order, quantum Hall-like insulating phase
(loop current order), next-nearest-neighbor (NNN) pairing order
and the density wave order, are numerically measured and two of
them (the plaquette/columnar VBS and density wave order) are
presented in Fig.~\ref{fig:OtherOrders} (the other two are
presented in supplemental material Sec. III~\cite{Suppl}). From
the extrapolations of structure factors, we conclude that none of
these operators develop long-range order near the DSM-FMI phase
transition.

\begin{figure}[tp!]
\centering
\includegraphics[width=0.9\columnwidth]{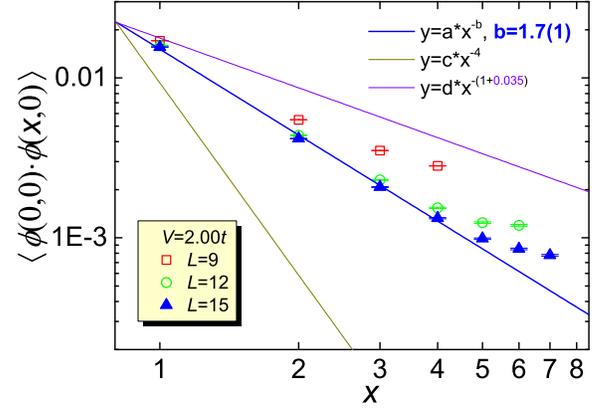}
\caption{\label{fig:Critial}(color online) Blue line: fit of the
spatial correlation of $O(6)$ order parameter $\boldsymbol{\phi}$
along $\mathbf{a}_1$ direction for $L=12,15$ systems as
$\langle\phi(0,0)\cdot\phi(x,0)\rangle$ at $V=V_c$. The obtained
anormalous dimension $\eta=0.7\pm0.1$. Dark green line:
$\frac{1}{x^{4}}$, the behavior of $O(6)$ correlation at $V=0$.
Violet line: $\frac{1}{x^{1.035}}$, the behavior of $O(6)$
correlation at the $(2+1)$D Wilson-Fisher $O(6)$ transition.}
\end{figure}

\textit{Continuous DSM-FMI phase transition}. The data of
excitation gaps and all possible order parameters reveal the
unusual mechanism of fermion mass generation without condensing
any fermion bilinear mass operator. To further explore the nature
of the DSM-FMI transition, we have also measured the 1st
derivative of ground state energy
$\langle\rho\rangle=\frac{1}{N_s}\frac{\partial{\langle\hat{H}\rangle}}{\partial{V}}=\frac{1}{N_s}\sum_{r}(c_{r1}^{\dagger}c_{r2}c_{r3}^{\dagger}c_{r4}+c_{r4}^{\dagger}c_{r3}c_{r2}^{\dagger}c_{r1})$.
The results are presented in Fig. S6 in supplemental
material~\cite{Suppl}. The converged $\langle\rho\rangle$ with
$L=15$ and $L=18$ changes continuously across the DSM-FMI phase
transition, indicating a continuous phase transition. Besides, we
have also measured the spatial correlation functions of $O(6)$
order parameter $\boldsymbol{\phi}$ along $\mathbf{a}_1$ direction
for $L=9,12,15$ at $V=V_c$, and the results are shown in
Fig.~\ref{fig:Critial}. In the log-log plot, convergence of the
slope for $L=12$ and $L=15$ can be seen. At the quantum critical
point, $\langle \phi(0,0) \cdot \phi(x,0) \rangle$ decays at
sufficiently long distances as $1/x^{1+\eta}$, where $\eta$ is the
anomalous dimension. Fit of the data gives $\eta=0.7\pm 0.1$. Such
anomalous dimension is much larger than that of the Wilson-Fisher
fixed point of $(2+1)$D $O(6)$ transition with $\eta=0.035$
obtained from $\epsilon$-expansion~\cite{Zinn-Justin2002}.
Also, spatial correlation of the O(6) order parameter of the
noninteracting Dirac fermions is shown in Fig.~\ref{fig:Critial},
which has a form of $1/x^{4}$.

\textit{Conclusions}. We find a continuous DSM-FMI transition
without any spontaneous symmetry breaking in a simple model of
four-flavor fermions with $SU(4)$ symmetry. The quantum critical
point at $V_c/t=2.00\pm0.05$ separate the gapless Dirac semimetal
from the featureless Mott insulator. Such new mechanism of mass
generation without fermion bilinear condensation is consistent
with previous studies from the lattice QCD
community~\cite{Ayyar2015a,Catterall2015,Ayyar2015b}.  More
interestingly, in our investigations, the excitation gaps and an exhaustive exclusion of symmetry breaking are for
the first time being directly accessed and a large anomalous
dimension $\eta$ at the DSM-FMI transition is revealed.

\textit{Acknowledgement}. We would like to thank S. Chandrasekharan, S. Catterall and H.-T. Ding for helpful discussions. The numerical calculations were carried out at the Physical Laboratory of High Performance Computing in RUC, the Center for Quantum Simulation Sciences in the Institute of Physics, Chinese Academy of Sciences, as well as the National Supercomputer Center in Tianjin (TianHe-1A) and GuangZhou (TianHe-2). H.Q.W., Y.Y.H., Z.Y.M. and Z.Y.L. acknowledge support from the National Natural Science Foundation of China (Grant Nos. 91421304, 11474356, 11421092 and 11574359). Z.Y.M. is also supported by the National Thousand-Young-Talents Program of China. Y.Z.Y. and C.X. are supported by the David and Lucile Packard Foundation and NSF Grant No. DMR-1151208.

\bibliography{SU4Conclusion}

\pagebreak

\onecolumngrid

\begin{center}
\textbf{\large Supplemental Material: Quantum critical point of fermion mass generation without spontaneous symmetry breaking}
\end{center}

\author{Yuan-Yao He}
\author{Han-Qing Wu}
\address{Department of Physics, Renmin University of China, Beijing 100872, China}
\author{Yi-Zhuang You}
\affiliation{Department of Physics, University of California,
Santa Barbara, California 93106, USA}
\author{Cenke Xu}
\affiliation{Department of Physics, University of California,
Santa Barbara, California 93106, USA}
\author{Zi Yang Meng}
\affiliation{Beijing National Laboratory
for Condensed Matter Physics, and Institute of Physics, Chinese
Academy of Sciences, Beijing 100190, China}
\author{Zhong-Yi Lu}
\address{Department of Physics, Renmin University of China, Beijing 100872, China}

\setcounter{equation}{0}
\setcounter{figure}{0}
\setcounter{table}{0}
\setcounter{page}{1}
\makeatletter
\renewcommand{\theequation}{S\arabic{equation}}
\renewcommand{\thefigure}{S\arabic{figure}}

\section{I. Projector QMC method and absence of sign problem}
\label{sec:QMCIntroduc}

\subsection{A. Projector QMC method}
Projector quantum Monte Carlo (PQMC) is the zero-temperature version of determinantal QMC algorithm~\cite{AssaadEvertz2008,Meng2010}. It obtains the ground-state observables by carrying out imaginary time evolution starting from a trial wavefunction that has overlap with the true many-body ground state. The ground-state expectation value of physical observable is calculated as follows,
\begin{equation}
\langle \hat{O}\rangle =\lim\limits_{\Theta\to +\infty}
\frac{\langle
\psi_T|e^{-\Theta\hat{H}/2}\hat{O}e^{-\Theta\hat{H}/2}|\psi_T\rangle}{\langle
\psi_T|e^{-\Theta \hat{H}}|\psi_T\rangle},
\label{eq:PQMC_Observable}
\end{equation}
where $|\psi_T\rangle$ is the trial wave function and $\Theta$ is the projection parameter. During the simulation, we choose $|\psi_T\rangle$ to be the ground state of the following single-particle Hamiltonian,
\begin{eqnarray}
\hat{H}_{\alpha}^0=-t\sum_{\langle lr\rangle}(-1)^{\alpha}c_{l\alpha}^{\dagger}c_{r\alpha} \ e^{2\pi i \frac{\Psi}{\Psi_0}} + h.c. ,
\end{eqnarray}
for every fermion flavor $\alpha$ and $\Psi_0=he/c$ is the flux quantum. The quantity $\Psi/\Psi_0=10^{-4}$ is chosen to lift the ground state degeneracy in $|\psi_T\rangle$ at the $K$ and $K'$ points where Dirac cones touch. For the model on honeycomb lattice, we perform QMC simulations on finite systems with linear size $L$ and lattice site $N_s=2L^2$ with periodic boundary conditions. To ensure that we have indeed projected out the ground state of the system, we choose $\Theta t=L+47$, in which the smallest $\Theta t=50$ is applied for $L=3$ systems and the largest $\Theta t=75$ is used for $L=18$ systems. The imaginary time discretization of $\Delta\tau t=0.05$ is applied for all the simulations.

\subsection{B. Absence of sign problem of the $SU(4)$ symmetric model}

To be able to perform QMC simulations, the absence of the minus-sign-problem is crucial, in this part, we discuss the reason that  the $SU(4)$ symmetric model in the main text is immune from the notorious minus-sign-problem.

The $SU(4)$ symmetric model reads,
\begin{eqnarray}
\label{eq:SU4SymModel}
\hat{H}=\underbrace{-t\sum_{\langle l,r \rangle\alpha}(-1)^{\alpha}(c_{l\alpha}^{\dagger}c_{r\alpha}+ c_{r\alpha}^{\dagger}c_{l\alpha})}_{\hat{H}_{\text{band}}}
{}+ \underbrace{V\sum_{r}(c_{r1}^{\dagger}c_{r2}c_{r3}^{\dagger}c_{r4} + c_{r4}^{\dagger}c_{r3}c_{r2}^{\dagger}c_{r1})}_{\hat{H}_{\text{int}}}.
\end{eqnarray}
First of all, we rewrite the interaction term $\hat{H}_{\text{int}}$ as
\begin{eqnarray}
\hat{H}_{\text{int}} &=& V\sum_r(c_{r1}^{\dagger}c_{r2}c_{r3}^{\dagger}c_{r4}+c_{r4}^{\dagger}c_{r3}c_{r2}^{\dagger}c_{r1}) \nonumber\\
&=&\frac{V}{2}\sum_r\Big[(\hat{D}_r)^2+(\hat{D}_r^{\dagger})^2\Big]\nonumber\\
&=&\frac{V}{4}\sum_r\Big[(\hat{D}_r+\hat{D}_r^{\dagger})^2+(\hat{D}_r-\hat{D}_r^{\dagger})^2\Big] \nonumber\\
&=&\frac{V}{4}\sum_r\Big[(\hat{D}_r+\hat{D}_r^{\dagger})^2 - (i\hat{D}_r-i\hat{D}_r^{\dagger})^2\Big],
\end{eqnarray}
where $\hat{D}_r=c_{r1}^{\dagger}c_{r2}+c_{r3}^{\dagger}c_{r4}$. In the last expression, we insert a imaginary unit $i$ to guarantee that the hopping term $(i\hat{D}_l-i\hat{D}_l^{\dagger})$ is Hermitian. We further split $\hat{H}_{\text{int}}$ into two parts, the kinetic part $\hat{H}_K$ (note $\hat{H}_K$ is different from $\hat{H}_{\text{band}}$) and the current part $\hat{H}_C$, as
\begin{eqnarray}
\hat{H}_K=\frac{V}{4}\sum_l(\hat{D}_l+\hat{D}_l^{\dagger})^2
\hspace{1.0cm} \hat{H}_C=-\frac{V}{4}\sum_l(i\hat{D}_l-i\hat{D}_l^{\dagger})^2
\end{eqnarray}
Then we can write the expression $e^{-\Theta\hat{H}}$ in the partition function $Z=\text{Tr}\{e^{-\Theta\hat{H}}\}$ as,
\begin{eqnarray}
\label{eq:Trotter}
e^{-\Theta\hat{H}}=e^{-\Theta(\hat{H}_{\text{band}}+\hat{H}_{\text{int}})}=(e^{-\Delta\tau(\hat{H}_{\text{band}}+\hat{H}_{\text{int})}})^M
      &&\approx (e^{-\Delta\tau\hat{H}_{\text{band}}}e^{-\Delta\tau\hat{H}_{\text{int}}})^M + \mathcal{O}[(\Delta\tau)^2] \nonumber\\
      &&\approx (e^{-\Delta\tau\hat{H}_{\text{band}}}e^{-\Delta\tau\hat{H}_K}e^{-\Delta\tau\hat{H}_C})^M + \mathcal{O}(\Delta\tau).
\end{eqnarray}
In this expression, the $\mathcal{O}[(\Delta\tau)^2]$ error in the first line comes from the commutator $[\hat{H}_0,\hat{H}_I]\ne 0$, while the $\mathcal{O}(\Delta\tau)$ error in the second line comes from the commutator $[\hat{H}_K,\hat{H}_C]\ne 0$. $\Delta\tau=\Theta/M$ is the discretization of imaginary time. Based on Eq.~(\ref{eq:Trotter}), we can now perform the Hubbard-Stratonovich (HS) transformation to decouple the interaction term.

To prove the absence of sign-problem for the $V>0$ case, we introduce a particle-hole transformation as follows
\begin{eqnarray}
\label{eq:ParticleHole1}
\text{(i)}\hspace{0.2cm}
\left\{\begin{array}{ll}
    c_{l1}\to           d_{l2}^{\dagger} \\
    c_{l1}^{\dagger}\to d_{l2}
          \end{array}
\right. \hspace{0.8cm}
\left\{\begin{array}{ll}
    c_{l2}\to           d_{l1}^{\dagger} \\
    c_{l2}^{\dagger}\to d_{l1}
          \end{array}
\right.  \hspace{2.0cm}
\left\{\begin{array}{ll}
    c_{l3}\to           d_{l4}^{\dagger} \\
    c_{l3}^{\dagger}\to d_{l4}
          \end{array}
\right.  \hspace{0.8cm}
\left\{\begin{array}{ll}
    c_{l4}\to           d_{l3}^{\dagger} \\
    c_{l4}^{\dagger}\to d_{l3}
          \end{array}
\right.
\end{eqnarray}
apply (i) to Eq.~(\ref{eq:SU4SymModel}), one can see that both $\hat{H}_{\text{band}}$ and $\hat{H}_{\text{int}}$ are invariant as
\begin{eqnarray}
+(c_{l1}^{\dagger}c_{r1} + c_{r1}^{\dagger}c_{l1}) && \xrightarrow{PH-(i)} +(d_{l2}d_{r2}^{\dagger} + d_{r2}d_{l2}^{\dagger})= -(d_{l2}^{\dagger}d_{r2} + d_{r2}^{\dagger}d_{l2})   \nonumber\\
-(c_{l2}^{\dagger}c_{r2} + c_{r2}^{\dagger}c_{l2}) && \xrightarrow{PH-(i)} -(d_{l1}d_{r1}^{\dagger} + d_{r1}d_{l1}^{\dagger})= +(d_{l1}^{\dagger}d_{r1} + d_{r1}^{\dagger}d_{l1})   \nonumber\\
c_{r1}^{\dagger}c_{r2}c_{r3}^{\dagger}c_{r4} + c_{r4}^{\dagger}c_{r3}c_{r2}^{\dagger}c_{r1} && \xrightarrow{PH-(i)}
d_{r2}d_{r1}^{\dagger}d_{r4}d_{r3}^{\dagger} + d_{r3}d_{r4}^{\dagger}d_{r1}d_{r2}^{\dagger}
= d_{r1}^{\dagger}d_{r2}d_{r3}^{\dagger}d_{r4} + d_{r4}^{\dagger}d_{r3}d_{r2}^{\dagger}d_{r1}.
\end{eqnarray}
Thus, the model Hamiltonian in Eq.~(\ref{eq:SU4SymModel}) is invariant under the particle-hole transformation (i) in Eq.~(\ref{eq:ParticleHole1}).

To prove the absence of sign-problem for the $V<0$ case, we introduce another particle-hole transformation as follows
\begin{eqnarray}
\label{eq:ParticleHole2}
\text{(ii)}\hspace{0.2cm}
\left\{\begin{array}{ll}
    c_{l1}\to           d_{l2}^{\dagger} \\
    c_{l1}^{\dagger}\to d_{l2}
          \end{array}
\right. \hspace{0.8cm}
\left\{\begin{array}{ll}
    c_{l2}\to           -d_{l1}^{\dagger} \\
    c_{l2}^{\dagger}\to -d_{l1}
          \end{array}
\right.  \hspace{2.0cm}
\left\{\begin{array}{ll}
    c_{l3}\to           d_{l4}^{\dagger} \\
    c_{l3}^{\dagger}\to d_{l4}
          \end{array}
\right.  \hspace{0.8cm}
\left\{\begin{array}{ll}
    c_{l4}\to           -d_{l3}^{\dagger} \\
    c_{l4}^{\dagger}\to -d_{l3}
          \end{array}
\right.
\end{eqnarray}
the particle-hole transformation in Eq.~(\ref{eq:ParticleHole2}) is different from the one in Eq.~(\ref{eq:ParticleHole1}) only up to simple sign. With (ii), $\hat{H}_{\text{band}}$ and $\hat{H}_{\text{int}}$ in Eq.~(\ref{eq:SU4SymModel}) become
\begin{eqnarray}
+(c_{l1}^{\dagger}c_{r1} + c_{r1}^{\dagger}c_{l1}) && \xrightarrow{PH-(ii)} +(d_{l2}d_{r2}^{\dagger} + d_{r2}d_{l2}^{\dagger})= -(d_{l2}^{\dagger}d_{r2} + d_{r2}^{\dagger}d_{l2})  \nonumber\\
-(c_{l2}^{\dagger}c_{r2} + c_{r2}^{\dagger}c_{l2}) && \xrightarrow{PH-(ii)} -(-1)^2(d_{l1}d_{r1}^{\dagger} + d_{r1}d_{l1}^{\dagger})= +(d_{l1}^{\dagger}d_{r1} + d_{r1}^{\dagger}d_{l1})  \nonumber\\
c_{r1}^{\dagger}c_{r2}c_{r3}^{\dagger}c_{r4} + c_{r4}^{\dagger}c_{r3}c_{r2}^{\dagger}c_{r1} && \xrightarrow{PH-(ii)}
(-1)^2(d_{r2}d_{r1}^{\dagger}d_{r4}d_{r3}^{\dagger} + d_{r3}d_{r4}^{\dagger}d_{r1}d_{r2}^{\dagger})
= d_{r1}^{\dagger}d_{r2}d_{r3}^{\dagger}d_{r4} + d_{r4}^{\dagger}d_{r3}d_{r2}^{\dagger}d_{r1}.
\end{eqnarray}
Thus, the model Hamiltonian in Eq.~(\ref{eq:SU4SymModel}) is also invariant under the particle-hole transformation (ii) defined in Eq.~(\ref{eq:ParticleHole2}).

Now we apply the HS transformation (i) for $V>0$. In the QMC, we decouple the interaction term $\hat{H}_{\text{int}}$ with the HS transformation of four-component Ising fields,
\begin{eqnarray}
\label{eq:HSTransf2}
&&\exp\left[-\Delta\tau\frac{V}{4}(\hat{D}_r+\hat{D}_r^{\dagger})^2\right]
= \frac{1}{4}\sum_{x=\pm1,\pm2}\gamma(x)e^{i\xi_V\eta(x)(\hat{D}_r+\hat{D}_r^{\dagger})} + \mathcal{O}\left[(\Delta\tau)^4\right] \nonumber\\
&&\exp\left[+\Delta\tau\frac{V}{4}(i\hat{D}_r-i\hat{D}_r^{\dagger})^2\right]
= \frac{1}{4}\sum_{x=\pm1,\pm2}\gamma(x)e^{\xi_V\eta(x)(i\hat{D}_r-i\hat{D}_r^{\dagger})} + \mathcal{O}\left[(\Delta\tau)^4\right],
\end{eqnarray}
where $\xi_V=\sqrt{\Delta\tau V/4}$, the auxiliary field $\{x=\pm1,\pm2\}$ live on the $(2+1)$D space-time lattice, and the coefficients $\gamma(x)$, $\eta(x)$ can be found in Refs.~\cite{AssaadEvertz2008,Meng2010,He2015a}. We can furthermore separate the hopping terms of $\alpha=1,2$ from those of $\alpha=3,4$,
\begin{eqnarray}
\label{eq:s8}
&&e^{i\xi_V\eta(x)(\hat{D}_r+\hat{D}_r^{\dagger})}=
e^{i\xi_V\eta(x)(c_{r1}^{\dagger}c_{r2}+c_{r2}^{\dagger}c_{r1})}
e^{i\xi_V\eta(x)(c_{r3}^{\dagger}c_{r4}+c_{r4}^{\dagger}c_{r3})} \nonumber\\
&&e^{\xi_V\eta(x)(i\hat{D}_r-i\hat{D}_r^{\dagger})}=
e^{\xi_V\eta(x)(ic_{r1}^{\dagger}c_{r2}-ic_{r2}^{\dagger}c_{r1})}
e^{\xi_V\eta(x)(ic_{r3}^{\dagger}c_{r4}-ic_{r4}^{\dagger}c_{r3})}.
\end{eqnarray}
We perform the particle-hole transformation (i) for the part with $\alpha=1,2$,
\begin{eqnarray}
&&c_{r1}^{\dagger}c_{r2}+c_{r2}^{\dagger}c_{r1} \xrightarrow{PH-(i)} d_{r2}d_{r1}^{\dagger}+d_{r1}d_{r2}^{\dagger} = -(d_{r1}^{\dagger}d_{r2}+d_{r2}^{\dagger}d_{r1})
\nonumber\\
&&c_{r1}^{\dagger}c_{r2}-c_{r2}^{\dagger}c_{r1} \xrightarrow{PH-(i)} d_{r2}d_{r1}^{\dagger}-d_{r1}d_{r2}^{\dagger} = -(d_{r1}^{\dagger}d_{r2}-d_{r2}^{\dagger}d_{r1})
\nonumber\\
&&\Rightarrow  \hspace{1.0cm} e^{i\xi_V\eta(x)(c_{r1}^{\dagger}c_{r2}+c_{r2}^{\dagger}c_{r1})} \xrightarrow{PH-(i)} e^{-i\xi_V\eta(x)(d_{r1}^{\dagger}d_{r2}+d_{r2}^{\dagger}d_{r1})}
\nonumber\\
&&\Rightarrow  \hspace{1.0cm} e^{\xi_V\eta(x)(ic_{r1}^{\dagger}c_{r2}-ic_{r2}^{\dagger}c_{r1})} \xrightarrow{PH-(i)} e^{-\xi_V\eta(x)(id_{r1}^{\dagger}d_{r2}-id_{r2}^{\dagger}d_{r1})}
\end{eqnarray}
one can observe that after the particle-hole transformation (i), the determinant related to the $\alpha=1,2$ flavors becomes complex-conjugate to the determinant related to the $\alpha=3,4$ flavor, and as the partition function is a product of the determinants for $\alpha=1,2$ and $\alpha=3,4$, the configurational weight for every HS field is positive definite.

For $V<0$, adopting the HS transformation as follows, we have
\begin{eqnarray}
\label{eq:HSTransf}
&&\exp\left[-\Delta\tau\frac{V}{4}(\hat{D}_r+\hat{D}_r^{\dagger})^2\right]
= \frac{1}{4}\sum_{x=\pm1,\pm2}\gamma(x)e^{\xi_V\eta(x)(\hat{D}_r+\hat{D}_r^{\dagger})} + \mathcal{O}\left[(\Delta\tau)^4\right]  \nonumber\\
&&\exp\left[+\Delta\tau\frac{V}{4}(i\hat{D}_r-i\hat{D}_r^{\dagger})^2\right]
= \frac{1}{4}\sum_{x=\pm1,\pm2}\gamma(x)e^{i\xi_V\eta(x)(i\hat{D}_r-i\hat{D}_r^{\dagger})} + \mathcal{O}\left[(\Delta\tau)^4\right],
\end{eqnarray}
where $\xi_V=\sqrt{-\Delta\tau V/4}$. After the HS transformations, we can again separate the hopping terms with $\alpha=1,2$ from those with $\alpha=3,4$
\begin{eqnarray}
&&e^{\xi_V\eta(x)(\hat{D}_r+\hat{D}_r^{\dagger})}=
e^{\xi_V\eta(x)(c_{r1}^{\dagger}c_{r2}+c_{r2}^{\dagger}c_{r1})}
e^{\xi_V\eta(x)(c_{r3}^{\dagger}c_{r4}+c_{r4}^{\dagger}c_{r3})}  \nonumber\\
&&e^{i\xi_V\eta(x)(i\hat{D}_r-i\hat{D}_r^{\dagger})}=
e^{i\xi_V\eta(x)(ic_{r1}^{\dagger}c_{r2}-ic_{r2}^{\dagger}c_{r1})}
e^{i\xi_V\eta(x)(ic_{r3}^{\dagger}c_{r4}-ic_{r4}^{\dagger}c_{r3})}.
\end{eqnarray}
We perform the particle-hole transformation (ii) for the part with $\alpha=1,2$,
\begin{eqnarray}
&&c_{r1}^{\dagger}c_{r2}+c_{r2}^{\dagger}c_{r1} \xrightarrow{PH-(ii)} -(d_{r2}d_{r1}^{\dagger}+d_{r1}d_{r2}^{\dagger})=d_{r1}^{\dagger}d_{r2}+d_{r2}^{\dagger}d_{r1}
\\ \nonumber
&&c_{r1}^{\dagger}c_{r2}-c_{r2}^{\dagger}c_{r1} \xrightarrow{PH-(ii)} -(d_{r2}d_{r1}^{\dagger}-d_{r1}d_{r2}^{\dagger})=d_{r1}^{\dagger}d_{r2}-d_{r2}^{\dagger}d_{r1}
\nonumber\\
&&\Rightarrow  \hspace{1.0cm} e^{\xi_V\eta(x)(c_{r1}^{\dagger}c_{r2}+c_{r2}^{\dagger}c_{r1})} \xrightarrow{PH-(ii)} e^{\xi_V\eta(x)(d_{r1}^{\dagger}d_{r2}+d_{r2}^{\dagger}d_{r1})}
\nonumber\\
&&\Rightarrow  \hspace{1.0cm} e^{i\xi_V\eta(x)(ic_{r1}^{\dagger}c_{r2}-ic_{r2}^{\dagger}c_{r1})} \xrightarrow{PH-(ii)} e^{i\xi_V\eta(x)(id_{r1}^{\dagger}d_{r2}-id_{r2}^{\dagger}d_{r1})}.
\end{eqnarray}
Again one sees that after the particle-hole transformation (ii), the determinant related to $\alpha=1,2$ is complex-conjugate to the determinant related to $\alpha=3,4$. Thus the $SU(4)$ symmetric model in Eq.~(\ref{eq:SU4SymModel}) with $V<0$ also has all its configurational weight positive definite, i.e., no sign-problem during the QMC simulation.

\section{II. Ground State Analysis}
\label{sec:GroundState}

First of all, for $V<0$ case, the exact ground state wavefunction of $\hat{H}_{\text{int}}$ in Eq.~(\ref{eq:SU4SymModel}) can be written as
\begin{eqnarray}
|\Psi_r\rangle = \frac{1}{\sqrt{2}}(c_{r1}^{\dagger}c_{r3}^{\dagger}+c_{r2}^{\dagger}c_{r4}^{\dagger})|0\rangle,
\end{eqnarray}
and we have $\hat{H}_{\text{int}}|\Psi_r\rangle=+V|\Psi_r\rangle$. Thus, at $V\to-\infty$, we can obtain the exact many-body wavefunction of the model Hamiltonian presented in Eq.~(\ref{eq:SU4SymModel}) as
\begin{eqnarray}
\label{eq:SU4ModelVle0}
|\Psi_g\rangle = \prod_r|\Psi_r\rangle=\prod_r\frac{1}{\sqrt{2}}(c_{r1}^{\dagger}c_{r3}^{\dagger}+c_{r2}^{\dagger}c_{r4}^{\dagger})|0\rangle
\hspace{0.4cm} \Rightarrow \hspace{0.4cm} \hat{H}_{\text{int}}|\Psi_g\rangle = +VN_s|\Psi_g\rangle.
\end{eqnarray}
We can explicitly observe that the state described by Eq.~(\ref{eq:SU4ModelVle0}) is a direct product state, which is free from fermion bilinear condensates. 

Alternatively, for $V>0$ case, the exact ground state wavefunction of $\hat{H}_{\text{int}}$ can be expressed as
\begin{eqnarray}
|\Psi_r\rangle = \frac{1}{\sqrt{2}}(c_{r1}^{\dagger}c_{r3}^{\dagger}-c_{r2}^{\dagger}c_{r4}^{\dagger})|0\rangle,
\end{eqnarray}
where we have $\hat{H}_{\text{int}}|\Psi_r\rangle=-V|\Psi_r\rangle$. Similarly, at $V\to+\infty$, we can obtain the exact many-body wavefunction of the model Hamiltonian presented in Eq.~(\ref{eq:SU4SymModel}) as
\begin{eqnarray}
\label{eq:SU4ModelVge0}
|\Psi_g\rangle = \prod_r|\Psi_r\rangle=\prod_r\frac{1}{\sqrt{2}}(c_{r1}^{\dagger}c_{r3}^{\dagger}-c_{r2}^{\dagger}c_{r4}^{\dagger})|0\rangle
\hspace{0.4cm} \Rightarrow \hspace{0.4cm} \hat{H}_{\text{int}}|\Psi_g\rangle = -VN_s|\Psi_g\rangle.
\end{eqnarray}
This is also a direct product state.

From the above analysis, we can see that at both limits $V\to+\infty$ and $V\to-\infty$, the model in Eq.~(\ref{eq:SU4SymModel}) possesses a unique ground state, which can be expressed as direct product state in real space. They both represent featureless Mott insulator, since both of them preserve all the lattice symmetry, $SU(4)$ symmetry and particle-hole symmetry, which together rule out all possible fermion bilinear mass terms. Furthermore, as shown in the main text, even away from the ideal $V\to\infty$ limit, the featureless Mott insulator is stable for a range of parameters, and there is no symmetry breaking in the ground state.

\section{III. Possible symmetry breaking orders}
\label{sec:AllOrders}

The model Hamiltonian in Eq.~(\ref{eq:SU4SymModel}) has $SU(4)$ symmetry, as well as discrete symmetries such as the particle-hole, translational, rotational and spatial inversion. In this part, we present an exhaustive analysis of all the possible long-range orders that breaks the symmetries of the model and generate fermion mass for Eq.~(\ref{eq:SU4SymModel}), and demonstrate our numerical results that all these symmetry breaking long-range orders are \textit{absent} in the DSM-FMI quantum phase transition.

\subsection{A. $SU(4)$ symmetry breaking orders}
\label{sec:SU4Breaking}

Without enlarging the unit-cell, there are 64 possible fermion bilinear terms (that are linearly independent) which explicitly break $SU(4)$ symmetry of the model Hamiltonian in Eq.~(\ref{eq:SU4SymModel}). They can be decomposed to irreducible representations of the $O(6)$ symmetry group as $64=1+6+15+20+15+6+1$, which stands for the representations of \textit{scalars} (1), \textit{vectors} (6), \textit{anti-symmetric tensors} (15), \textit{symmetric tensors} (20), \textit{pseudo anti-symmetric tensors} (15), \textit{pseudo vectors} (6), \textit{pseudo scalars} (1). Among them, the vector, anti-symmetric tensor, pseudo vector and pseudo scalar representations are on-site fermion bilinear terms, while the rest of the representations are inter-site fermion bilinear terms.

\textit{Vector and Pseudo-Vector}.
The $O(6)$ vector and $O(6)$ pseudo-vector are actually the vectors $\boldsymbol{\phi}_l$ and $\boldsymbol{\psi}_l$ defined in the main text. Here, we list them again,
\begin{eqnarray}
\left\{\begin{array}{llllll}
    \phi_{l1}=\text{Re}(c_{l1}^{\dagger}c_{l4}+c_{l3}^{\dagger}c_{l2}) \\
    \phi_{l2}=\text{Re}(c_{l1}^{\dagger}c_{l3}^{\dagger}+c_{l2}c_{l4}) \\
    \phi_{l3}=\text{Re}(c_{l1}^{\dagger}c_{l2}-c_{l3}^{\dagger}c_{l4}) \\
    \phi_{l4}=\text{Im}(c_{l1}^{\dagger}c_{l4}-c_{l3}^{\dagger}c_{l2}) \\
    \phi_{l5}=\text{Im}(c_{l1}^{\dagger}c_{l3}^{\dagger}-c_{l2}c_{l4}) \\
    \phi_{l6}=\text{Im}(c_{l1}^{\dagger}c_{l2}+c_{l3}^{\dagger}c_{l4})
          \end{array}
\right.  \hspace{1.5cm}
\left\{\begin{array}{llllll}
    \psi_{l1}=\text{Im}(c_{l1}^{\dagger}c_{l4}+c_{l3}^{\dagger}c_{l2}) \\
    \psi_{l2}=\text{Im}(c_{l1}^{\dagger}c_{l3}^{\dagger}+c_{l2}c_{l4}) \\
    \psi_{l3}=\text{Im}(c_{l1}^{\dagger}c_{l2}-c_{l3}^{\dagger}c_{l4}) \\
    \psi_{l4}=\text{Re}(c_{l1}^{\dagger}c_{l4}-c_{l3}^{\dagger}c_{l2}) \\
    \psi_{l5}=\text{Re}(c_{l1}^{\dagger}c_{l3}^{\dagger}-c_{l2}c_{l4}) \\
    \psi_{l6}=\text{Re}(c_{l1}^{\dagger}c_{l2}+c_{l3}^{\dagger}c_{l4})
          \end{array}
\right. .
\end{eqnarray}
The $SU(4)\simeq SO(6)$ symmetry rotates the six components in $\boldsymbol{\phi}_l$ (and $\boldsymbol{\psi}_l$) to one another and these six orders are degenerate. This is rather like the spin $SU(2)$ symmetric Hubbard model, in which the spin $SU(2)\simeq SO(3)$ rotates the three components of spinor $\boldsymbol{S}=(S_x,S_y,S_z)$. Thus, both the real space correlations and structure factors of the six components in $\boldsymbol{\phi}_l$ are exactly the same, which is also the case for $\boldsymbol{\psi}_l$. Thus, we only need measure the correlations of one component in both $\boldsymbol{\phi}_l$ and $\boldsymbol{\psi}_l$, in principle. To improve the data quality, in the QMC simulation, we measure the results for all six components and then take the average. After that, the extrapolation of their structure factors $P(\boldsymbol{\Gamma})/N$ and $Q(\boldsymbol{\Gamma})/N$, corresponding to $\boldsymbol{\phi}_l$ and $\boldsymbol{\psi}_l$, are shown in Fig. 2 in the main text. The results explicitly show that there is no such long-range orders in the $SU(4)$ symmetric model.

\begin{figure}[tp!]
\centering
\includegraphics[width=\columnwidth]{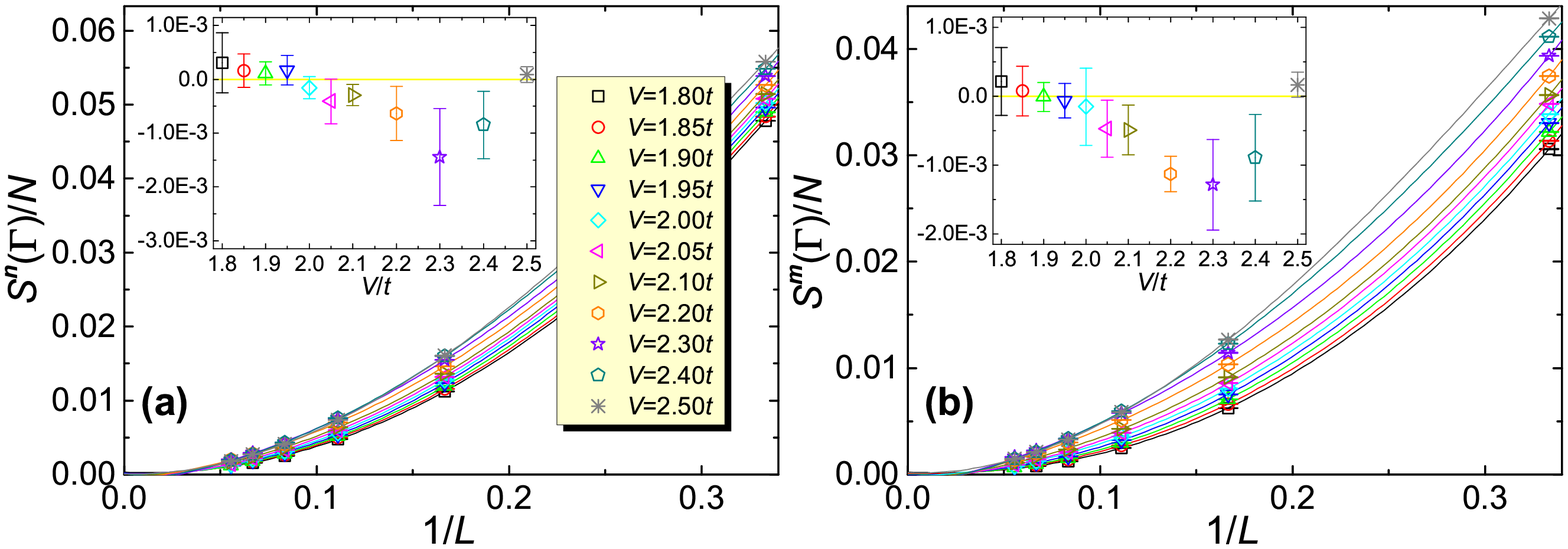}
\caption{\label{fig:OtherO6Order}(color online) Extrapolation of structure factors (a) $S^{\boldsymbol{n}}(\boldsymbol{\Gamma})/N$ for $\mathbf{n}_{l}$ orders and (b) $S^{\boldsymbol{m}}(\boldsymbol{\Gamma})/N$ for $\mathbf{m}_{l}$ orders over the inverse system size $1/L$ by cubic polynomials. The insets shows the extrapolation values at $L\to\infty$. Results show both of these $O(6)$ orders are absent across the DSM-FMI phase transition.}
\end{figure}

Despite the above definition of the \textit{Vector and Pseudo-Vector}, we can also combine $\boldsymbol{\phi}_l$ and $\boldsymbol{\psi}_l$ to define $6$ complex order parameters as,
\begin{eqnarray}
\label{eq:S23}
n_{l1}&=c_{l1}^{\dagger}c_{l2},\qquad
n_{l2}&=c_{l3}^{\dagger}c_{l4},\\
n_{l3}&=c_{l1}^{\dagger}c_{l3}^{\dagger},\qquad
n_{l4}&=c_{l4}c_{l2},\\
n_{l5}&=c_{l1}^{\dagger}c_{l4},\qquad
n_{l6}&=c_{l2}c_{l3}^{\dagger},
\end{eqnarray}
which are actually the six possible bilinear term of the $SU(4)$ spinor $\xi_r=(c_{r1}^{\dagger},c_{r2},c_{r3}^{\dagger},c_{r4})^{\text{T}}$. By the $SU(4)\simeq SO(6)$ symmetry, the orders corresponding to the six components vector $\mathbf{n}_{l}=(n_{l1},n_{l2},n_{l3},n_{l4},n_{l5},n_{l6})$ are degenerate and they have exactly the same correlations. Furthermore, there are also non-vanishing off-diagonal correlations as $\langle n_{l1}n_{l2} \rangle$, $\langle n_{l3}n_{l4} \rangle$ and $\langle n_{l5}n_{l6} \rangle$, which we denote as $\mathbf{m}_{l}=(m_{l1},m_{l2},m_{l3})$ orders. The $SU(4)$ symmetry also guarantees that these three orders are exactly degenerate. The manner of applying $\boldsymbol{n}_{l}$ and $\boldsymbol{m}_{l}$ to define the \textit{Vector and Pseudo-Vector} orders has actually been adopted in Ref.~\onlinecite{Ayyar2015a,Ayyar2015b}. To make a detailed comparison, we have also measured the correlations for the $\mathbf{n}_{l}$ and $\mathbf{m}_{l}$ orders. The extrapolations of their corresponding structure factors divided by $N$ are presented in Fig.~\ref{fig:OtherO6Order}, which also suggests that both the $\mathbf{n}_{l}$ and $\mathbf{m}_{l}$ long-range orders are absent, in consistent with the results presented in Fig. 2 in main text.

\textit{Pseudo Scalars and Anti-Symmetric Tensors}.
The pseudo scalar order parameter is just the CDW order as,
\begin{eqnarray}
\Delta_{l}^{CDW} = (-1)^l(c_{l1}^{\dagger}c_{l1} - c_{l2}^{\dagger}c_{l2} + c_{l3}^{\dagger}c_{l3} - c_{l4}^{\dagger}c_{l4}).
\end{eqnarray}
Here we choose to define the notion of "charge" in the $SU(4)$ spinor basis, meaning that the $U(1)$ transformation generated by the charge operator is $\xi_{l}\to e^{i\theta}\xi_{l}$. The anti-symmetric tensors are simply $SU(4)$ generators, there are 15 of them. We do not need to check each of them, because they are all related by symmetries. Here, we choose the generators of three $U(1)$ subgroups in $SU(4)$ symmetry group, which corresponds to various SDW orders as,
\begin{eqnarray}
g_{l}^{(1)} &=& (-1)^l(c_{l1}^{\dagger}c_{l1} + c_{l2}^{\dagger}c_{l2} + c_{l3}^{\dagger}c_{l3} + c_{l4}^{\dagger}c_{l4} - 2) \nonumber\\
g_{l}^{(2)} &=& (-1)^l(c_{l1}^{\dagger}c_{l1} + c_{l2}^{\dagger}c_{l2} - c_{l3}^{\dagger}c_{l3} - c_{l4}^{\dagger}c_{l4}) \nonumber\\
g_{l}^{(3)} &=& (-1)^l(c_{l1}^{\dagger}c_{l1} - c_{l2}^{\dagger}c_{l2} - c_{l3}^{\dagger}c_{l3} + c_{l4}^{\dagger}c_{l4}).
\end{eqnarray}
We can combine the CDW and SDW orders and enumerate all the density orders into the four fermion flavor channels as,
\begin{eqnarray}
\rho_{l1} &= (-1)^l(c_{l1}^{\dagger}c_{l1} - 1/2),\qquad
\rho_{l2} &= (-1)^l(c_{l2}^{\dagger}c_{l2} - 1/2),\\
\rho_{l3} &= (-1)^l(c_{l3}^{\dagger}c_{l3} - 1/2),\qquad
\rho_{l4} &= (-1)^l(c_{l4}^{\dagger}c_{l4} - 1/2).
\end{eqnarray}
The $-1/2$ is to ensure that $\langle \rho_{l\alpha} \rangle=0$ if there is no density order. The correlations of these four components are exactly the same. If one wants to exclude all the long-range density orders, it's sufficient to choose arbitrary one of $\rho_{l\alpha}$ to check whether its correlation is short-ranged. For example, we choose the first one and define the correlation as
\begin{eqnarray}
\label{eq:DensityWave}
D_{lr}^{DW} = \langle \rho_{l1}\rho_{r1} \rangle = (-1)^{l+r}\langle (c_{l1}^{\dagger}c_{l1} - 1/2)(c_{r1}^{\dagger}c_{r1} - 1/2) \rangle
\end{eqnarray}
The corresponding structure factor can also be defined and measured. The extrapolation of the structure factor divided by $N$, corresponding to the correlation in Eq.~(\ref{eq:DensityWave}), is shown in Fig. 4(b) of the main text. From the results, the correlation in Eq.~(\ref{eq:DensityWave}) is indeed short-ranged and all the long-range density orders can be excluded.

\begin{figure}[tp!]
\centering
\includegraphics[width=\columnwidth]{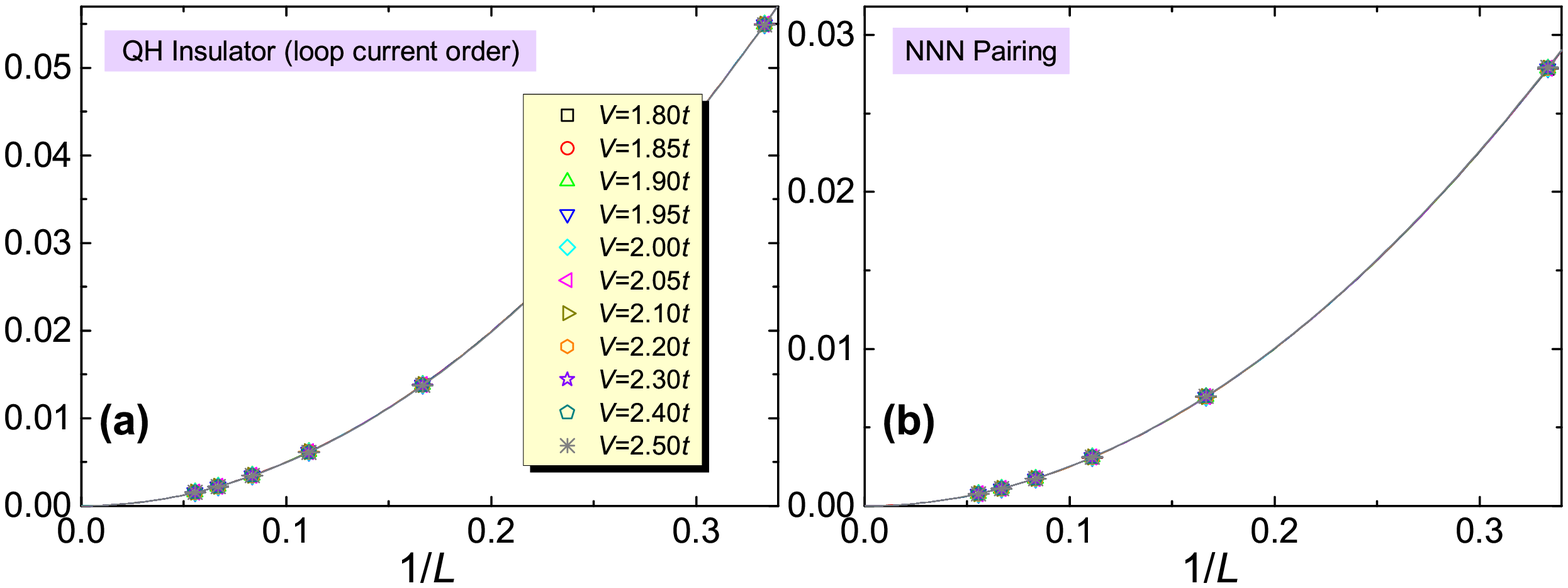}
\caption{\label{fig:OtherAllOrder}(color online) Extrapolation of structure factors for (a) quantum-Hall like loop current order and (b) next-nearest-neighbor pairing order over inverse system size $1/L$ by cubic polynomials, across the DSM-FMI phase transition. The results explicitly shows that none of these two long-range orders exist near the DSM-FMI phase transition. }
\end{figure}

\textit{Scalars and Pseudo Anti-Symmetric Tensors}.
The scalar order parameter is just the quantum spin Hall (QSH) order defined as,
\begin{eqnarray}
\label{eq:QSH}
\Delta_{lr}^{QSH} =(-1)^l(ic_{l1}^{\dagger}c_{r1} + ic_{l2}^{\dagger}c_{r2} + ic_{l3}^{\dagger}c_{r3} + ic_{l4}^{\dagger}c_{r4} + h.c.),
\end{eqnarray}
where $l,r$ are the sites connected by a next-nearest-neighbor (NNN) bond. Note that $l$ and $r$ always belong to the same sublattice, therefore it makes sense to define the sublattice sign just by $(-1)^l$. By definition $\Delta_{lr}^{QSH}$ is anti-symmetric under the exchange $l\leftrightarrow r$ as $\Delta_{lr}^{QSH}=-\Delta_{rl}^{QSH}$, which is a common feature of all the inter-site fermion bilinear terms we considered in the following. There are 15 pseudo anti-symmetric tensors. Again we choose the simplest ones to check. Here we choose different kinds of QSH-like orders,
\begin{eqnarray}
\Delta_{lr}^{(1)} &=&(-1)^l( ic_{l1}^{\dagger}c_{r1} + ic_{l2}^{\dagger}c_{r2} - ic_{l3}^{\dagger}c_{r3} - ic_{l4}^{\dagger}c_{r4} + h.c.),  \\
\Delta_{lr}^{(2)} &=& (-1)^l(ic_{l1}^{\dagger}c_{r1} - ic_{l2}^{\dagger}c_{r2} - ic_{l3}^{\dagger}c_{r3} + ic_{l4}^{\dagger}c_{r4} + h.c.),  \\
\Delta_{lr}^{(3)} &=& (-1)^l(ic_{l1}^{\dagger}c_{r1} - ic_{l2}^{\dagger}c_{r2} + ic_{l3}^{\dagger}c_{r3} - ic_{l4}^{\dagger}c_{r4} + h.c.).
\end{eqnarray}
Combining all the QSH-like order, we can just enumerate all the NNN imaginary hoppings as,
\begin{eqnarray}
u_{lr}^{(1)} &= (-1)^l(ic_{l1}^{\dagger}c_{r1} + h.c.),\qquad
u_{lr}^{(2)} &= (-1)^l(ic_{l2}^{\dagger}c_{r2} + h.c.),\\
u_{lr}^{(3)} &= (-1)^l(ic_{l3}^{\dagger}c_{r3} + h.c.),\qquad
u_{lr}^{(4)} &= (-1)^l(ic_{l4}^{\dagger}c_{r4} + h.c.),
\end{eqnarray}
where $lr$ is connected by a NNN bond. To make sure none of them are ordered, it will be sufficient to choose arbitrary one of them and check the correlation is short-ranged. For example, we choose the first one and define the correlation function as,
\begin{eqnarray}
\label{eq:AllQHOrder}
D_{ij,lr}^{QH}= \langle u_{ij}^{(1)}u_{lr}^{(1)} \rangle = - (-1)^{i+l}\langle (c_{i1}^{\dagger}c_{j1}-c_{j1}^{\dagger}c_{i1}) (c_{l1}^{\dagger}c_{r1}-c_{r1}^{\dagger}c_{l1}) \rangle,
\end{eqnarray}
where $ij$ and $lr$ are sites connected by NNN bonds and their bonds are of parallel orientations. The extrapolation of structure factor corresponding to the correlation in Eq.~(\ref{eq:AllQHOrder}) is presented in Fig.~\ref{fig:OtherAllOrder} (a). The results support the absence of all long-range quantum-Hall like orders.

\textit{Symmetric Tensors}.
There are 20 symmetric tensors. We picked eight of them which can be combined to the following four (complex) NNN pairing orders as,
\begin{eqnarray}
\label{eq:PairingOrder}
\omega_{lr}^{(1)} &= c_{l1}^{\dagger}c_{r1}^{\dagger},\qquad\omega_{lr}^{(2)} &= c_{l2}c_{r2},\\
\omega_{lr}^{(3)} &= c_{l3}^{\dagger}c_{r3}^{\dagger},\qquad\omega_{lr}^{(4)} &= c_{l4}c_{r4},
\end{eqnarray}
where $ij$ is NNN bond. All the other orders in the $20$ symmetric tensors correspond to nearest-neighbor (NN) pairing or more distant parings. First, the NN pairing order only breaks the $C_3$ symmetry of the system, which can only shift the position of the Dirac points in the energy spectrum when the order is weak. Thus, close to the DSM-FMI phase transition, if such order steps in, it can't open the gap and generate fermion mass. Taking that point into account, we simply neglect such pairing order. Second, considering the local interaction in the model of Eq.~(\ref{eq:SU4SymModel}), more distant pairing order than that of NNN pairing is rather unlikely to exist. Based on these considerations, we only concentrate on the NNN pairing orders defined in Eq.~(\ref{eq:PairingOrder}). To make sure none of them are ordered, it will be sufficient to choose arbitrary one of them and check the correlation is short-ranged. For example, we choose the first one and define the correlation function as,
\begin{eqnarray}
\label{eq:AllPairingOrder}
D_{ij,lr}^{SC}= \langle \omega_{ij}^{(1)}(\omega_{lr}^{(1)})^{\dagger} \rangle = \langle c_{i1}^{\dagger}c_{j1}^{\dagger}c_{r1}c_{l1} \rangle,
\end{eqnarray}
where $ij$ and $lr$ are sites connected by NNN bonds and their bonds are of parallel orientations. The extrapolation of structure factor corresponding to the correlation in Eq.~(\ref{eq:AllPairingOrder}) is presented in Fig.~\ref{fig:OtherAllOrder} (b). The results shows that all the NNN pairing long-range orders are absent.

\subsection{B. Discrete symmetry breaking orders}
\label{sec:DiscreteBreaking}

All the $SU(4)$ symmetry breaking orders aforementioned preserve the transitional symmetry. In this session, we concentrate on translational symmetry breaking order, which is the valence bond solid (VBS) order. On honeycomb lattice, there are three different kinds of VBS orders, i.e. the dimer VBS, plaquette VBS and columnar VBS~\cite{Lang2013,Zhou2015}. 
Dimer VBS only breaks $C_3$ symmetry, and similar to the nearest-neighbor pairing order, it only moves the Dirac points instead of opening single-particle gap. Thus, we will not discuss dimer VBS and only focus on the plaquette and columnar VBS orders, which breaks the transitional symmetry and can generate fermion mass. They can be sorted into the four orders parameters as,
\begin{eqnarray}
v_{lr}^{(1)} = c_{l1}^{\dagger}c_{r1} + c_{r1}^{\dagger}c_{l1}  \hspace{1.0cm}
v_{lr}^{(2)} = c_{l2}^{\dagger}c_{r2} + c_{r2}^{\dagger}c_{l2}  \hspace{1.0cm}
v_{lr}^{(3)} = c_{l3}^{\dagger}c_{r3} + c_{r3}^{\dagger}c_{l3}  \hspace{1.0cm}
v_{lr}^{(4)} = c_{l4}^{\dagger}c_{r4} + c_{r4}^{\dagger}c_{l4},
\end{eqnarray}
where $lr$ is connected by a nearest-neighbor bond. To make sure none of them are ordered, it will be sufficient to choose arbitrary one of them and check whether the correlation is short-ranged. For example, we choose the first one and define the correlation function as,
\begin{eqnarray}
\label{eq:AllVBSOrder}
D_{ij,lr}^{VBS}= \langle v_{ij}^{(1)}v_{lr}^{(1)} \rangle - \langle v_{ij}^{(1)} \rangle\langle v_{lr}^{(1)} \rangle =  \langle (c_{i1}^{\dagger}c_{j1}+c_{j1}^{\dagger}c_{i1}) (c_{l1}^{\dagger}c_{r1}+c_{r1}^{\dagger}c_{l1}) \rangle
- \langle c_{i1}^{\dagger}c_{j1}+c_{j1}^{\dagger}c_{i1} \rangle  \langle c_{l1}^{\dagger}c_{r1}+c_{r1}^{\dagger}c_{l1} \rangle,
\end{eqnarray}
where $ij$ and $lr$ are sites connected by nearest-neighbor bonds and their bonds are of parallel orientations. The extrapolation of structure factor corresponding to the correlation in Eq.~(\ref{eq:AllVBSOrder}) is presented in Fig. 4 (a) in the main text. The results explicitly exclude both plaquette and columnar VBS long-range orders.

\begin{figure}[tp!]
\centering
\includegraphics[width=\columnwidth]{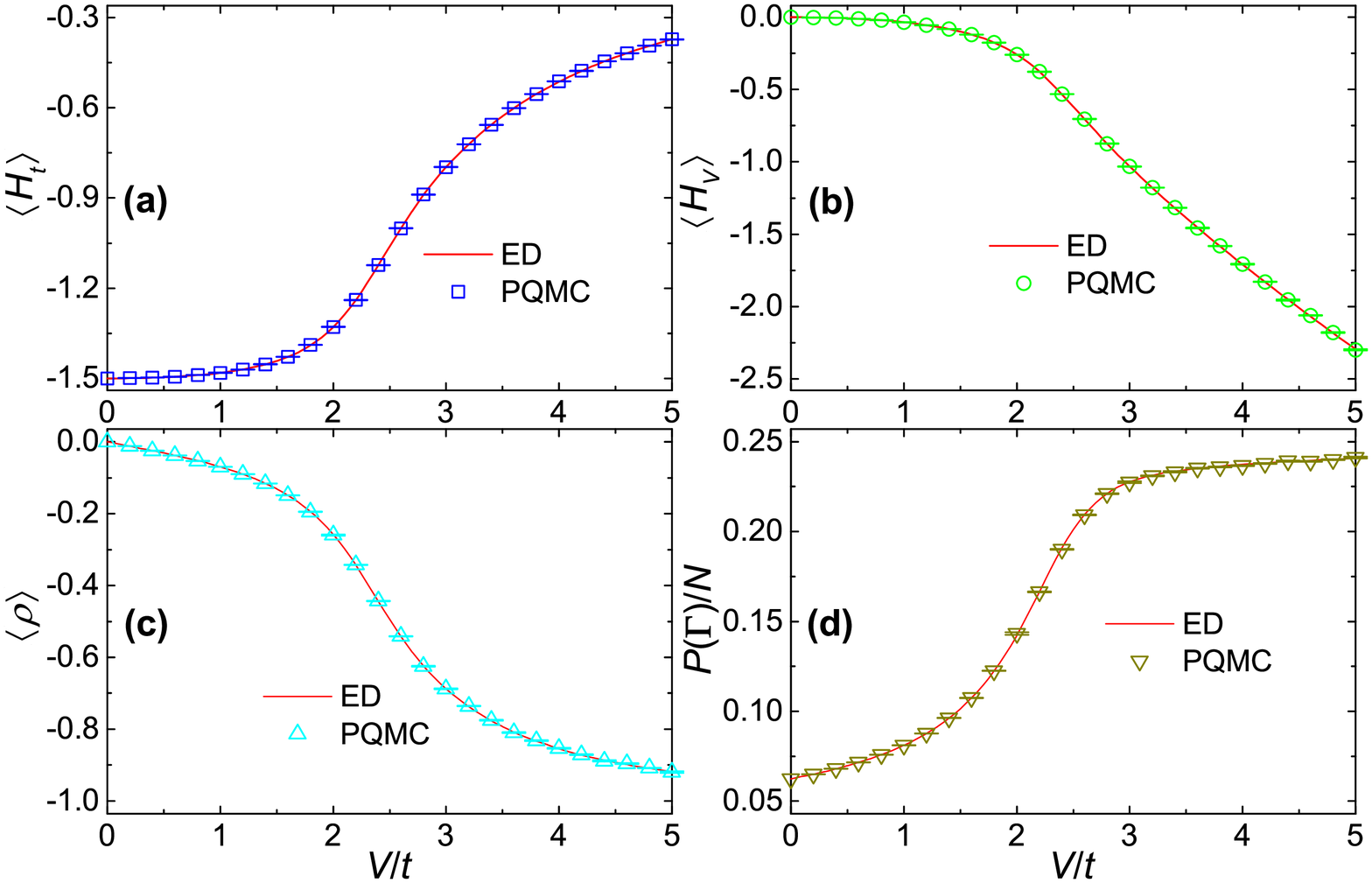}
\caption{\label{fig:EDQMCCompare}(color online) The comparisons between QMC and ED results on a $2\times2$ unit cells system for (a) energy density $\langle H_t \rangle$, (b) energy density $\langle H_V \rangle$, (c) effective order parameter $\langle\hat{\rho}\rangle$ and (d) structure factor $P(\boldsymbol{\Gamma})/N$. Perfect consistency between QMC and ED can be observed.}
\end{figure}

\section{IV. Sanity check of QMC simulations}
\label{sec:NumericalCheck}

We have also carried out sanity check to make sure that the QMC simulation results of the $SU(4)$ symmetric model in Eq.~(\ref{eq:SU4SymModel}) are correct in all respects. The first check is the comparison between QMC and exact diagonalization (ED) on a $2\times2$ unit cells system with 8 sites. Second, we have numerically verified the $SU(4)$ symmetry of the model in Eq.~(\ref{eq:SU4SymModel}). Third, we present raw data of  dynamic correlations to show that the extracted excitation gaps from QMC simulations are of high quality.

\subsection{A. Comparison between QMC and ED}

We measure the energy densities ($\langle H_t \rangle = \langle\hat{H}_{\text{band}}\rangle/N_s$ and $\langle H_V \rangle = \langle\hat{H}_{\text{int}}\rangle/N_s$) and the structure factors ($P(\mathbf{\Gamma})/N$) for the model of Eq.~(\ref{eq:SU4SymModel}) to compare the QMC and ED results. We have also measured the effective order parameter for the $SU(4)$ symmetric model defined as,
\begin{eqnarray}
\label{eq:EffectiveOrder}
\hat{\rho} = \frac{1}{N_s}\sum_{r}(c_{r1}^{\dagger}c_{r2}c_{r3}^{\dagger}c_{r4} + c_{r4}^{\dagger}c_{r3}c_{r2}^{\dagger}c_{r1}).
\end{eqnarray}
According to Hellmann-Feynman theorem, the expectation value of $\hat{\rho}$ is actually the fist-order derivative of total energy per site over the model parameter $V$. So we can use this quantity to determine whether the $V$-driven phase transition is of first-order or continuous, depending on whether $\langle\hat{\rho}\rangle$ is diverging or continuous around the phase transition point. As for structure factor, $P(\boldsymbol{\Gamma})/N$, it is the $O(6)$ vector order $\phi_l$ defined in Eq.~(3) in main text. During the QMC simulations of the $2\times2$ systems, we choose a special set of parameters $\Theta t=35,\Delta\tau t=0.01$ due to the small system size.

The comparisons of QMC and ED results are presented in Fig.~\ref{fig:EDQMCCompare}. We can observe that the results from QMC and from ED are well consistent with each other.

\subsection{B. Numerical verification of $SU(4)$ symmetry}

\begin{figure}[tp!]
\centering
\includegraphics[width=\columnwidth]{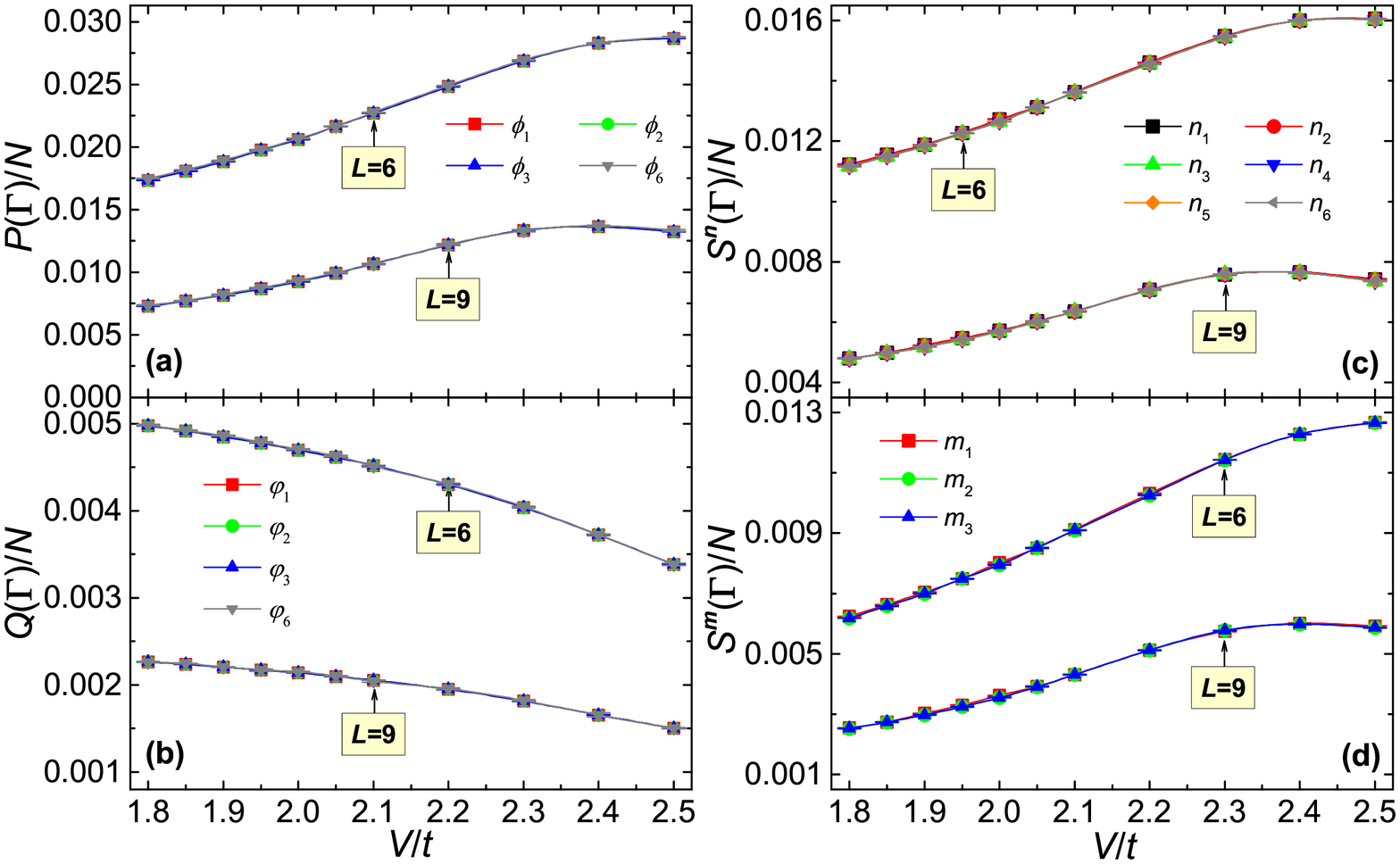}
\caption{\label{fig:SU4NumCheck}(color online) Numerical verification of $SU(4)$ symmetry of the model Hamiltonian in Eq.~(\ref{eq:SU4SymModel}). (a), (b) are the structure factors $P(\boldsymbol{\Gamma})/N$ for $\phi_l$ orders and $Q(\boldsymbol{\Gamma})/N$ for $\psi_l$ orders, in which we only present the results for $\alpha=1,2,3,6$ components. (c), (d) are the structure factors $S^n(\boldsymbol{\Gamma})/N$ for $\mathbf{n}_{l}$ orders and $S^m(\boldsymbol{\Gamma})/N$ for $\mathbf{m}_{l}$ orders.}
\end{figure}

The $SU(4)$ symmetry of the model Hamiltonian in Eq.~(\ref{eq:SU4SymModel}) guarantees that the structure factors for all the six components in $\boldsymbol{\phi}_l$ are exactly the same for finite-size systems (no spontaneous symmetry breaking), and the same holds for the six components in $\boldsymbol{\psi}_l$ as well. Moreover, during QMC implementation, we observe that after applying the Wick theorem, the equalities of $\langle\phi_{l1}\phi_{r1}\rangle\equiv\langle\phi_{l4}\phi_{r4}\rangle$ and $\langle\phi_{l2}\phi_{r2}\rangle\equiv\langle\phi_{l5}\phi_{r5}\rangle$ hold at the operator level. This suggests that we only need to compare $\phi_{l\alpha},\alpha=1,2,3,6$ components, the same holds for $\psi_l$ orders. Similarly, due to the $SU(4)$ symmetry, the structure factors of every component $S^n(\boldsymbol{\Gamma})/N$ for the $\mathbf{n}_{l}$ order should be exactly equal for finite-size systems, while the structure factors of every component $S^m(\boldsymbol{\Gamma})/N$ for the $\mathbf{m}_{l}$ order are also exactly the same.

To numerically verify the $SU(4)$ symmetry, we compare the results of structure factors of 4 components in $\phi_l$ vector and $\psi_l$ vector of $L=6,9$ systems, respectively. The results are shown in Fig.~\ref{fig:SU4NumCheck} (a), (b). Alternatively, we also compare the results of structure factors of 6 components in $\mathbf{n}_{l}$ vector and 3 components in $\mathbf{m}_l$ vector of $L=6,9$ systems. The results are shown in Fig.~\ref{fig:SU4NumCheck} (c), (d). The results are well consistent with our expectations, so within errorbars the $SU(4)$ symmetry of the model Hamiltonian in Eq.~(\ref{eq:SU4SymModel}) is indeed confirmed by our QMC simulations.

\subsection{C. Raw data of dynamic correlation functions}

In Fig. (3) of the main text, we have presented the data of excitation gaps in both fermionic and bosonic channels, from which we extrapolate the gap values in thermodynamic limit. Here, we show some raw data of both dynamic single-particle Green's function $G(\mathbf{K},\tau)$ and the dynamic correlation function $P(\boldsymbol{\Gamma},\tau)$ for the $\boldsymbol{\phi}_l$ orders, to demonstrate that the excitation gaps in Fig. (3) in the main text are extracted from good quality imaginary-time displaced data. $G(\mathbf{K},\tau)$ is defined in Eq. (4) in the main text, and $P(\boldsymbol{\Gamma},\tau)$ is defined in Eq. (5) in the main text.

\begin{figure}[tp!]
\centering
\includegraphics[width=\columnwidth]{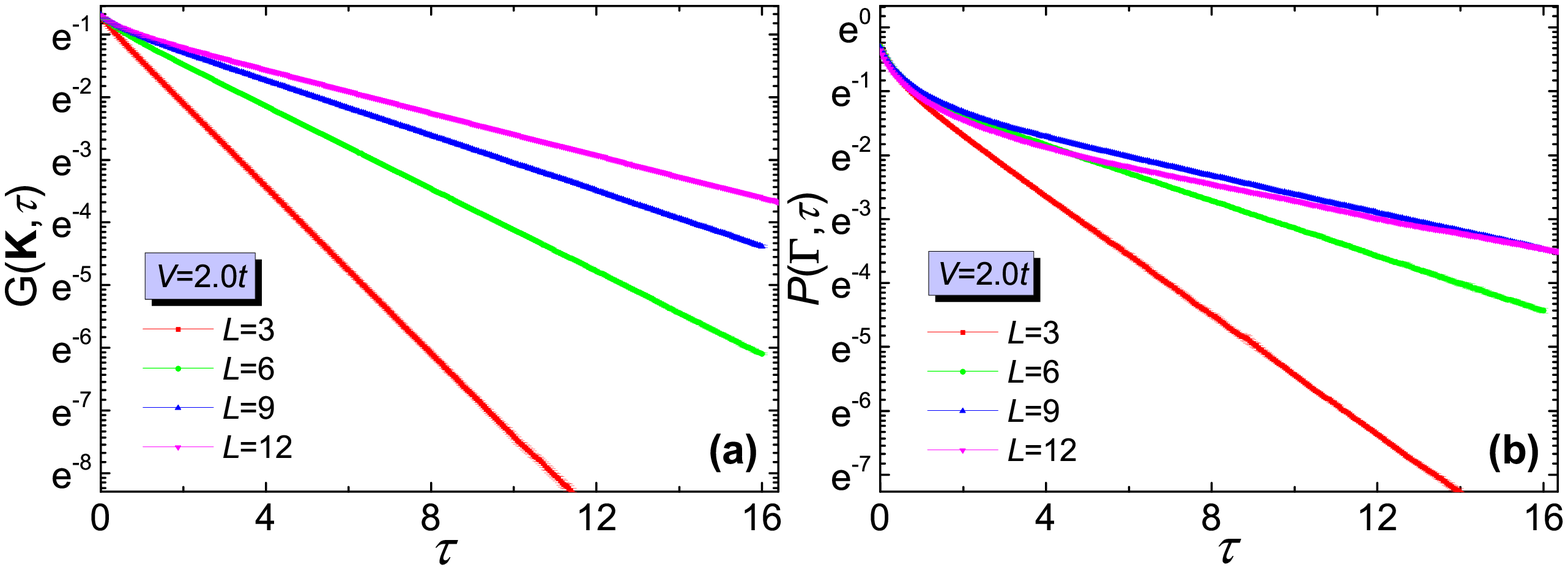}
\caption{\label{fig:DynamicData}(color online) The raw data of dynamic correlation function for the model in Eq.~(\ref{eq:SU4SymModel}) with $V/t=2.0$ in $L=3,6,9,12$ systems. (a) $G(\mathbf{K},\tau)$ and (b) $P(\boldsymbol{\Gamma},\tau)$  in semilogarithmic coordinate. Perfectly linear lines with $\tau$ can be observed, indicating high quality of the raw data.}
\end{figure}

The results of $G(\mathbf{K},\tau)$ and $P(\boldsymbol{\Gamma},\tau)$ with increasing $\tau$ in semilogarithmic coordinate are shown in Fig.~\ref{fig:DynamicData} (a) and (b). We can observe that the lines in Fig.~\ref{fig:DynamicData} are prefectly linear at long time (large $\tau$). The slopes of these linear lines are the corresponding values of excitation gaps.

\begin{figure}[tp!]
\centering
\includegraphics[width=0.7\columnwidth]{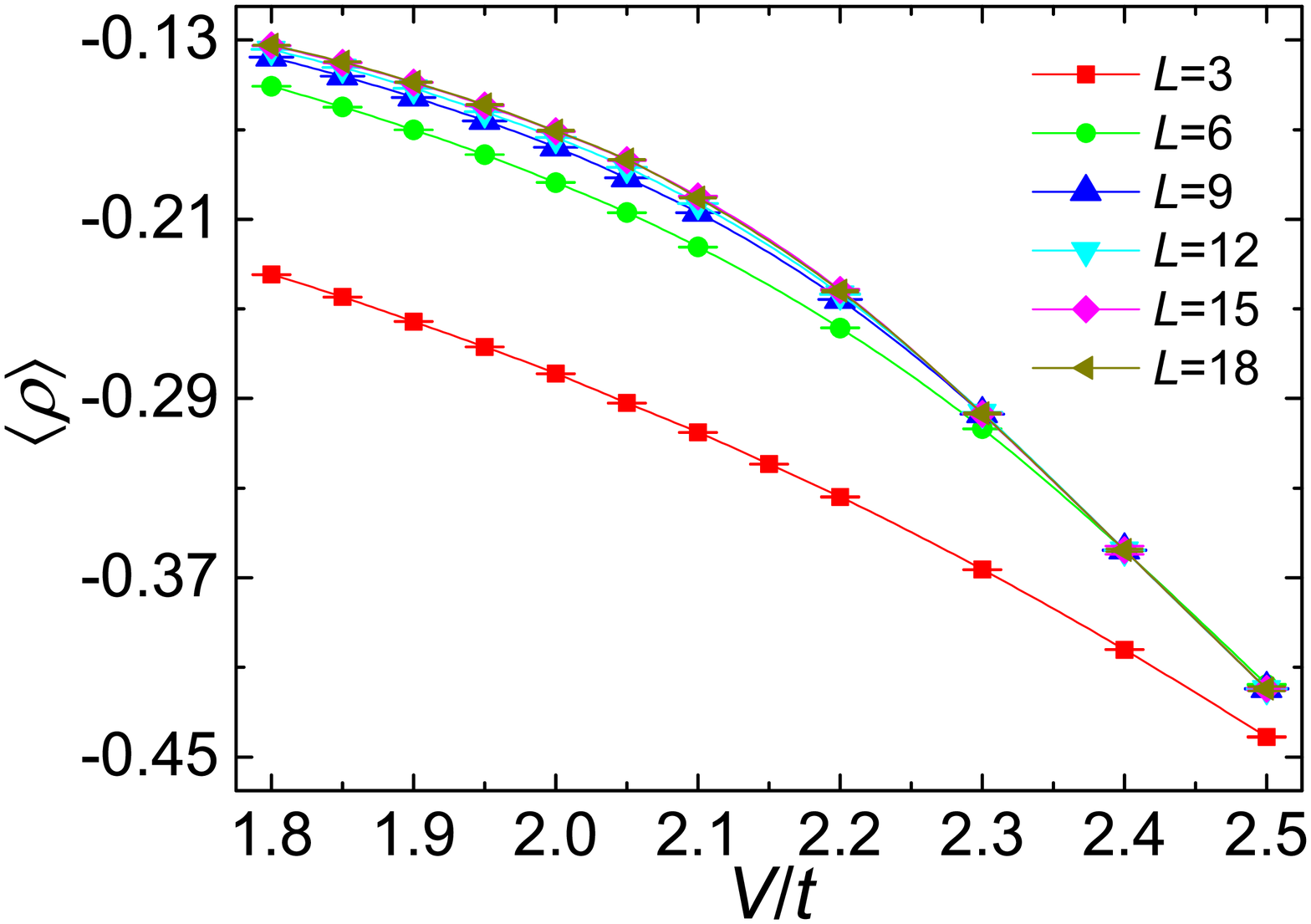}
\caption{\label{fig:FirstOrderDerivative}(color online) QMC results of $\langle\rho\rangle$ for $L=3,6,9,12,15, 18$ systems with $1.8t\le V\le2.5t$ around the DSM-FMI phase transition. Converged results can be observed for $L=15,18$ systems, which indicates that $\langle\rho\rangle$ changes continuously across the DSM-FMI phase transition. }
\end{figure}

\section{V. Continuous DSM-FMI phase transition}
\label{sec:ContinuousPhaseTran}

In the main text, we present numerical results supporting a direct DSM-FMI phase transition, and in the whole $V$ region there is no spontaneous symmetry breaking. However, since there is no nonzero local order parameter for the model Hamiltonian to distinguish the Dirac Semimetal and the featureless Mott insulator, it's difficult to determine whether the direct DSM-FMI phase transition is of first-order or continuous. Here we present the numerical results of $\langle\rho\rangle$ defined in Eq.~(\ref{eq:EffectiveOrder}), which suggests that the DSM-FMI phase transition is continuous. As mentioned above, $\langle\rho\rangle$ is actually the fist-order derivative of the total energy per site over the model parameter $V$ as $\langle\rho\rangle=\frac{1}{N_s}\frac{\partial{\langle\hat{H}\rangle}}{\partial{V}}$. At zero temperature and thermodynamic limit, if $\langle\rho\rangle$ diverges at the quantum phase transition point, then the phase transition is of first order. Otherwise if it's continuous, it suggests a continuous phase transition.

Numerical results of $\langle\rho\rangle$ for $L=3,6,9,12,15, 18$ systems with $1.8t\le V\le2.5t$, which is around the DSM-FMI phase transition is shown in Fig.~\ref{fig:FirstOrderDerivative}. As system sizes increase, we can indeed obtain converged results of $\langle\rho\rangle$. As shown in Fig.~\ref{fig:FirstOrderDerivative}, we can observe that $\langle\rho\rangle$ has very little changes (about $5\times10^{-4}$) from $L=15$ to $L=18$, indicating that $\langle\rho\rangle$ has almost reached its thermodynamic values. The converged $\langle\rho\rangle$ changes continuously in the chosen region around the DSM-FMI phase transition, suggesting a continuous quantum phase transition.


\end{document}